\documentclass{elsart}
\newcommand{\hs}{\hspace*{0.5cm}}

\newcommand{\be}{\begin{equation}}
\newcommand{\ee}{\end{equation}}
\newcommand{\bea}{\begin{eqnarray}}
\newcommand{\eea}{\end{eqnarray}}
\newcommand{\nn}{\nonumber}
\newcommand{\crn}{\nonumber \\}

\newcommand{\al}{\alpha}
\newcommand{\la}{\lambda}
\newcommand{\bet}{\beta}
\newcommand{\ga}{\gamma}

\newcommand{\om}{\omega}

\newcommand{\fr}{\frac}
\newcommand{\bc}{\begin{center}}
\newcommand{\ec}{\end{center}}

\newcommand{\La}{\Lambda}

\newcommand {\ba}{\begin{array}}
\newcommand {\ea}{\end{array}}
\newcommand{\ben}{\begin{enumerate}}
\newcommand{\een}{\end{enumerate}}

\begin{document}
\runauthor{Dong, Huong, Marcos and Long}
\begin{frontmatter}
\title{Supersymmetric economical 3-3-1 model}
\author[Hanoi]{P. V. Dong\thanksref{D}}
\author[Hanoi]{D. T. Huong\thanksref{H}}
\author[Brazil]{ M. C. Rodriguez\thanksref{M}}
\author[Hanoi]{H. N. Long\thanksref{L}}

\address[Hanoi]{Institute of Physics, VAST, P.O. Box 429, Bo Ho, Hanoi
10000, Vietnam}
\address[Brazil]{Funda\c{c}\~{a}o Universidade Federal do Rio Grande-FURG,
 Departamento de F\'\i sica,
 Av. It\'alia, km 8, Campus Carreiros,
 96201-900 Rio Grande, RS,  Brazil}
\thanks[D]{Email: pvdong@iop.vast.ac.vn}
\thanks[H]{Email: dthuong@iop.vast.ac.vn}
\thanks[M]{Email: mcrodriguez@einstein.fisica.furg.br}
\thanks[L]{Email: hnlong@iop.vast.ac.vn}

\begin{abstract}
The supersymmetric extension of the economical 3-3-1 model is
presented. The constraint equations and the gauge boson
identification establish a relation  between the vacuum
expectation values (VEVs) at the top and bottom elements of the
Higgs triplet $\chi$ and its supersymmetric counterpart
$\chi^\prime$. Because of this relation, the exact diagonalization
of neutral gauge boson sector has been performed. The gauge bosons
and their associated Goldstone ones mix in the same way as in
non-supersymmetric version. This is also correct in the case of
gauginos. The eigenvalues and eigenstates in the Higgs sector are
derived. The model contains a heavy neutral Higgs boson with  mass
equal to those of the neutral non-Hermitian gauge boson $X^0$  and
a charged scalar with mass  equal to those of the  $W$ boson in
the standard model, i. e. $ m_{\varrho_1} = m_W$.  This result is
in good agreement
 with the present estimation: $m_{H^\pm} > 79.3$ GeV, CL=
95 \%. We also show that the boson sector and the fermion sector
gain masses in the same way as in the non-supersymmetric case.

\end{abstract}
\end{frontmatter}

\section{\label{intro}Introduction}
Recent neutrino  experimental results~\cite{superk,kam,sno}
establish the fact that neutrinos have masses and the standard
model (SM) must be extended. The generation of neutrino masses
is an important issue in any realistic
extension of the SM. In general, the values of these masses (of the
order of, or less than, 1 eV) that are needed to explain all neutrino
oscillation data are not enough to put strong
constraints on model building. It means that several models can induce neutrino
masses and mixing compatible with experimental data. So, instead of
proposing models built just to explain the neutrino properties, it is more
useful to consider what are the neutrino masses that are predicted in any
particular model which has motivation other than the explanation of neutrino
physics.

The SM is exceedingly successful in describing leptons,
quarks and their interactions. Nevertheless, the SM is
not considered as the ultimate theory since neither the fundamental
parameters, masses and couplings, nor the symmetry pattern are
predicted. These elements are merely built into the model. Likewise,
the spontaneous electroweak symmetry breaking is simply parametrized
by a single Higgs doublet field.

Even though many aspects of the SM are experimentally supported to
a very accuracy, the embedding of the model into a more general
framework is to be expected. The argument is closely connected to
the mechanism of the electroweak symmetry breaking. If the Higgs
boson is light, the SM can naturally be embedded in a grand
unified theory, the so called GUT. The large energy gap between
the low electroweak scale and the high GUT scale can be stabilized
by supersymmetry. Supersymmetry actually provides the link between
the experimentally explored interactions at electroweak energy
scales and physics at scales close to the Planck scale $M_{pl}
\approx 10^{19}$ GeV where gravity is important.

On the other hand, the possibility of a gauge symmetry based on
the following symmetry $\mathrm{SU}(3)_{C} \otimes
\mathrm{SU}(3)_{L} \otimes \mathrm{U}(1)_{X}$
(3-3-1)~\cite{ppf,flt,331rh} is particularly interesting, because
it explains some fundamental questions that are eluded in the SM.
The main motivations to study this kind of model are:
\begin{enumerate}
\item The family number must be multiple of three;
\item It explains why $\sin^{2} \theta_{W}<\frac{1}{4}$ is observed;
\item It solves the strong CP problem;
\item It is the simplest model that includes bileptons of both types: scalar
and vectors ones;
\item The model has several sources of CP violation.
\end{enumerate}

In one of the 3-3-1 models \cite{331rh} which is anomaly free, the
particle content is given by \footnote{In this article the
notation is slightly different from those in Ref. \cite{s331r}.}
\bea L_{aL} &=& \left(\nu_{a}, l_{a}, \nu_{a}^c \right)^T_L \sim (
{\bf 3}, -1/3),\ l_{aR}\sim ( {\bf 1}, -1),\ a = 1, 2, 3, \crn
Q_{1L}&=&\left( u_{1},  d_{1}, u' \right)^T_L\sim \left( {\bf
3},1/3\right),\ Q_{\al L}=\left(d_{\al},  -u_{\al},  d'_{\al}
\right)^T_L\sim ( {\bf 3}^*,0),\ \al=2,3,\crn u_{i R}&\sim&\left(
{\bf 1},2/3\right),\ d_{i R} \sim \left( {\bf 1},-1/3\right),\
i=1,2,3, \crn u'_{R}&\sim& \left( {\bf 1},2/3\right),\ d'_{\al R}
\sim \left( {\bf 1},-1/3\right), \nn\eea where the values in the
parentheses denote quantum numbers based on the
$\left(\mbox{SU}(3)_L,\mbox{U}(1)_X\right)$ symmetry. The exotic
quarks $u'$ and $d'_\al$ in this case take the same electric
charges as of the usual quarks, i.e., $q_{u'}=2/3$,
$q_{d'_\al}=-1/3$. The spontaneous symmetry breaking in this model
is achieved by two Higgs scalar triplets only \be \chi
=\left(\chi^0_1, \chi^-, \chi^0_2 \right)^T \sim \left( {\bf
3},-1/3\right),\ \rho=\left(\rho^+_1, \rho^0, \rho^+_2\right)^T
\sim \left( {\bf 3},2/3\right)\ee with all the neutral components
$\chi^0_{1}$, $\chi^0_2$ and $\rho^0$ developing the vacuum
expectation values.

Such a scalar sector is minimal, and therefore it has been called
the economical 3-3-1 model~\cite{ponce,haihiggs}. In a series of
papers, we have developed and proved that this non-supersymmetric
version is consistent, realistic and very rich in physics. Let us
remind some steps in the development: The general Higgs sector is
very simple (economical) and consists of three physical scalars
(two neutral and one charged) and eight Goldstone bosons - the
needed number for massive gauge ones~\cite{higgseconom}. In
Refs.\cite{dlhh,dls1},  we have  shown that the model under the
consideration is realistic, by the mean that, at the  one-loop
level, all fermions gain consistent masses. In addition, the model
contains a Majoron associated with $\chi^0_1$ responsible for the
Majorana masses of neutrinos.

Supersymmetry (SUSY) that transforms fermions into bosons and vice
versa is a leading candidate for physics beyond the
SM~\cite{susy}. The existence of such a non-trivial extension is
highly constrained by theoretical principles. One of the
motivations for supersymmetry is that it can help to understand
the hierarchy problem. The supersymmetric version of the 3-3-1
model with right-handed neutrinos~\cite{331rh} has already been
constructed in Refs.~\cite{s331r,hrl}. The neutrino masses in this
case were studied in Ref.\cite{marcos}, and  the proton
instability was considered in Ref~\cite{pallong}.

It was shown that~\cite{sidm331}, the 3-3-1 models are the first
gauge ones containing the candidates for self-interaction dark
matter (SIDM)~\cite{SIDM} with the condition given by Spergel and
Steinhardt~\cite{ss}. It was shown that~\cite{higgseconom} the
economical 3-3-1 model does not furnish any candidate for SIDM.
This directly relates to the scalar sector in which a significant
number of fields and couplings is reduced (to compare, see Ref.
~\cite{hrl}). With the larger field content, the supersymmetric
version of the model is expected to provide candidates for dark
matter, particularly for the SIDM.

The aim of this work is to study the supersymmetric version of the
economical 3-3-1 model.

The outline of this paper is as follows. In Sec.~\ref{parcontent}
we present a fermion and scalar content in the  supersymmetric
economical 3-3-1 model. The supersymmetric Lagrangian and breaking
are given in Sec.~\ref{sec:mass}. In Sections \ref{gaugeboson},
\ref {leptonquark}
 and \ref{sec:massspectrum} we deal with the gauge
boson, fermion  and Higgs sectors of the model. Finally, we
summarize our results and make conclusions in the last section -
Sec. \ref{concl}.

\section{\label{parcontent}Particle content
of supersymmetric economical 3-3-1 model}

To proceed further, the necessary features of the supersymmetric
economical 3-3-1 model \cite{haihiggs} will be presented. The
superfield content in this paper is defined in a standard way as
follows \be \widehat{F}= (\widetilde{F}, F),\hs \widehat{S} = (S,
\widetilde{S}),\hs \widehat{V}= (\lambda,V), \ee where the
components $F$, $S$ and $V$ stand for the fermion, scalar and
vector fields of the economical 3-3-1 model while their
superpartners are denoted as $\widetilde{F}$, $\widetilde{S}$ and
$\lambda$, respectively \cite{susy,s331r}.

The superfields for the leptons under the 3-3-1 gauge group
transform as
\begin{equation}
\widehat{L}_{a L}=\left(\widehat{\nu}_{a}, \widehat{l}_{a},
\widehat{\nu}^c_{a}\right)^T_{L} \sim ( {\bf 1},{\bf 3},-1/3),\hs
  \widehat {l}^{c}_{a L} \sim ({\bf 1},{\bf 1},1),\label{l2}
\end{equation} where $\widehat{\nu}^c_L=(\widehat{\nu}_R)^c$ and $a=1,2,3$
is a generation index.

It is worth mentioning that, in the economical version the first
generation of quarks should be different from others \cite{dlhh}.
The superfields for the left-handed quarks of the first generation
are in triplets \be \widehat Q_{1L}= \left(\widehat { u}_1,\
                        \widehat {d}_1,\
                        \widehat {u}^\prime
 \right)^T_L \sim ( {\bf 3},{\bf 3},1/3),\label{quarks3}\ee
where the right-handed singlet counterparts are given by\be
\widehat {u}^{c}_{1L},\ \widehat { u}^{ \prime c}_{L} \sim ( {\bf
3^{*}},{\bf 1},-2/3),\hs \widehat {d}^{c}_{1L} \sim ( {\bf
3^{*}},{\bf 1},1/3 ). \label{l5} \ee Conversely, the superfields
for the last two generations transform as antitriplets
\begin{equation}
\begin{array}{ccc}
 \widehat{Q}_{\alpha L} = \left(\widehat{d}_{\alpha}, - \widehat{u}_{\alpha},
 \widehat{d^\prime}_{\alpha}\right)^T_{L} \sim (
 {\bf 3},{\bf 3^*},0), \hs \al=2,3, \label{l3}
\end{array}
\end{equation}
where the right-handed counterparts are in singlets
\begin{equation}
\widehat{u}^{c}_{\alpha L} \sim \left( {\bf 3^{*}},{\bf 1},-2/3
\right),\hs \widehat{d}^{c}_{\alpha L},\ \widehat{d}^{\prime
c}_{\alpha L} \sim \left( {\bf 3^{*}},{\bf 1},1/3 \right).
\label{l4}
\end{equation}

The primes superscript on usual quark types ($u'$ with the
electric charge $q_{u'}=2/3$ and $d'$ with $q_{d'}=-1/3$) indicate
that those quarks are exotic ones. The mentioned fermion content,
which belongs to that of the 3-3-1 model with right-handed
neutrinos \cite{331rh,haihiggs} is, of course,  free from anomaly.

The two superfields $\widehat{\chi}$ and $\widehat {\rho} $ are at
least introduced to span the scalar sector of the economical 3-3-1
model \cite{higgseconom}: \bea \widehat{\chi}&=& \left (
\widehat{\chi}^0_1, \widehat{\chi}^-, \widehat{\chi}^0_2
\right)^T\sim (1,3,-1/3), \label{l7}\\
\widehat{\rho}&=& \left (\widehat{\rho}^+_1, \widehat{\rho}^0,
\widehat{\rho}^+_2\right)^T \sim  (1,3,2/3). \label{l8} \eea To
cancel the chiral anomalies of Higgsino sector, the two extra
superfields $\widehat{\chi}^\prime$ and $\widehat {\rho}^\prime $
must be added as follows \bea \widehat{\chi}^\prime&=& \left
(\widehat{\chi}^{\prime 0}_1, \widehat{\chi}^{\prime
+},\widehat{\chi}^{\prime 0}_2 \right)^T\sim ( 1,3^*,1/3),
\label{l9}\\
\widehat{\rho}^\prime &=& \left (\widehat{\rho}^{\prime -}_1,
  \widehat{\rho}^{\prime 0},  \widehat{\rho}^{\prime -}_2
\right)^T\sim (1,3^*,-2/3). \label{l10} \eea

In this model, the $ \mathrm{SU}(3)_L \otimes \mathrm{U}(1)_X$
gauge group is broken via two steps:
 \be \mathrm{SU}(3)_L \otimes
\mathrm{U}(1)_X \stackrel{w,w'}{\longrightarrow}\ \mathrm{SU}(2)_L
\otimes \mathrm{U}(1)_Y\stackrel{v,v',u,u'}{\longrightarrow}
\mathrm{U}(1)_{Q},\label{stages}\ee where the VEVs are defined by
\bea
 \sqrt{2} \langle\chi\rangle^T &=& \left(u, 0, w\right), \hs \sqrt{2}
 \langle\chi^\prime\rangle^T = \left(u^\prime,  0,
 w^\prime\right),\\
\sqrt{2}  \langle\rho\rangle^T &=& \left( 0, v, 0 \right), \hs
\sqrt{2} \langle\rho^\prime\rangle^T = \left( 0, v^\prime,  0
\right).\eea The VEVs $w$ and $w^\prime$ are responsible for the
first step of the symmetry breaking while $u,\ u^\prime$ and $v,\
v^\prime$ are for the second one. Therefore, they have to satisfy
the constraints:
 \be
 u,\ u^\prime,\ v,\ v^\prime
\ll w,\ w^\prime. \label{contraint}\ee

The vector superfields $\widehat{V}_c$, $\widehat{V}$ and
$\widehat{V}^\prime$ containing the usual gauge bosons are,
respectively, associated with the $\mathrm{SU}(3)_C$,
$\mathrm{SU}(3)_L$ and $\mathrm{U}(1)_X $ group factors. The
colour and flavour vector superfields have expansions in the
Gell-Mann matrix bases $T^a=\lambda^a/2$ $(a=1,2,...,8)$ as
follows\bea \widehat{V}_c &=& \fr{1}{2}\lambda^a
\widehat{V}_{ca},\hs
\widehat{\overline{V}}_c=-\fr{1}{2}\lambda^{a*}
\widehat{V}_{ca};\hs \widehat{V} = \fr{1}{2}\lambda^a
\widehat{V}_{a},\hs \widehat{\overline{V}}=-\fr{1}{2}\lambda^{a*}
\widehat{V}_{a},\eea where an overbar $^-$ indicates complex
conjugation. For the vector superfield associated with
$\mathrm{U}(1)_X$, we normalize as follows \be X \hat{V}'= (XT^9)
\hat{B}, \hs T^9\equiv\fr{1}{\sqrt{6}}\mathrm{diag}(1,1,1).\ee In
the following, we are denoting the gluons by $g^a$ and their
respective gluino partners by $\lambda^a_{c}$, with $a=1,
\ldots,8$. In the electroweak sector, $V^a$ and $B$ stand for the
$\mathrm{SU}(3)_{L}$ and $\mathrm{U}(1)_{X}$ gauge bosons with
their gaugino partners $\lambda^a_{V}$ and $\lambda_{B}$,
respectively.

\section {\label{sec:mass}Lagrangian}
With the superfields  given above, we can now construct the
supersymmetric economical 3-3-1 model containing the Lagrangians:
$\mathcal{L}_{susy}+\mathcal{L}_{soft}$, where the first term is
supersymmetric part, whereas the last term breaks explicitly the
supersymmetry.

\subsection{\label{suplagran}Supersymmetric Lagrangian}

The supersymmetric Lagrangian can be decomposed into four relevant
parts
\begin{eqnarray} \mathcal{L}_{susy} &=& {\mathcal L}_{gauge}
+{\mathcal L}_{lepton} +{\mathcal L}_{quark} +{\mathcal
L}_{scalar}. \label{lg2}
\end{eqnarray}
The first term has the form
\begin{eqnarray}
{\mathcal L}_{gauge} &=& \frac{1}{4} \int
d^{2}\theta\;\mathcal{W}_{ca} \mathcal{W}_{ca}+ \frac{1}{4} \int
d^{2}\theta\;\mathcal {W}_a \mathcal{W}_a+ \frac{1}{4} \int
d^{2}\theta
\mathcal{W}^{\prime}\mathcal{W}^{ \prime} \nonumber \\
&& + \frac{1}{4} \int
d^{2}\bar{\theta}\;\overline{\mathcal{W}}_{ca}\overline{\mathcal{W}}_{ca}+
\frac{1}{4} \int d^{2}\bar{\theta}\;\overline{\mathcal{W}}_a
\overline{\mathcal{W}}_a+ \frac{1}{4} \int  d^{2}\bar{\theta}
\overline{\mathcal{W}}^{ \prime}\overline{\mathcal{W}}^{
\prime},\label{gaug}
\end{eqnarray}
where the chiral superfields $\mathcal{W}_{c}$, $\mathcal{W}$ and
$\mathcal{W}^{ \prime}$ are defined by
\begin{eqnarray}
\mathcal{W}_{c\zeta}&=&- \frac{1}{8g_s} \bar{D} \bar{D} e^{-2g_s
\hat{V}_{c}}
D_{\zeta} e^{2g_s \hat{V}_{c}},\nonumber \\
\mathcal{W}_{\zeta}&=&- \frac{1}{8g} \bar{D} \bar{D} e^{-2g \hat{V}}
D_{\zeta} e^{2g \hat{V}}, \nonumber \\
\mathcal{W}^{\prime}_{\zeta}&=&- \frac{1}{4} \bar{D} \bar{D}
D_{\zeta} \hat{V}^{\prime}, \,\ \zeta=1,2, \label{cforca}
\end{eqnarray}
with the gauge couplings $g_{s}$, $g$ and $g^{\prime}$ respective
to $\mathrm{SU}(3)_C$, $\mathrm{SU}(3)_L$ and $\mathrm{U}(1)_X$.
The $D_{\zeta}$ and $\bar{D}_{\dot{\zeta}}$ are the chiral
covariant derivatives of SUSY algebra as presented in \cite{susy}.

The second and third terms are given by \begin{eqnarray} {\mathcal
L}_{lepton} &=& \int d^{4}\theta \left[ \hat{\bar{L}}_{a
L}e^{2(g\hat{V}- \frac{g^\prime}{3}\hat{V}^\prime)}\hat{L}_{a L}
+\hat{\bar{l}}^c_{a L}e^{2g^\prime \hat{V}^\prime}\hat{l}^c_{a L}
\right], \label{lg3}
\end{eqnarray}
 and
\begin{eqnarray}
{\mathcal L}_{quark} &=& \int d^{4}\theta \left[ \hat{\bar{Q}}_{1
L}e^{{2(g_s\hat{V}_c+g\hat{V}+\frac{g^\prime}{3}
\hat{V}^\prime)}}\hat{Q}_{1 L} + \hat{\bar{Q}}_{\alpha
L}e^{{2(g_s\hat{V}_c+g\hat{\bar{V}})}}\hat{Q}_{\alpha L}
\right.\crn  &&+\hat{\bar{u}}^c_{iL}e^{2(g_s
\hat{\bar{V}}_c-\frac{2g^\prime}{3}\hat{V}^\prime) }\hat{u}^c_{iL}
+\hat{\bar{d}}^c_{iL}e^{2(g_s
\hat{\bar{V}}_c+\frac{g^\prime}{3}\hat{V}^\prime)}\hat{d}^c_{iL}
\nonumber \\ &&+\left. \hat{\bar{u}}^{\prime c}_{L}e^{2(g_s
\hat{\bar{V}}_c-\frac{2g^\prime}{3}\hat{V}^\prime
)}\hat{u}^{\prime c}_{L}+\hat{\bar{d}}^{\prime c}_{\alpha
L}e^{2(g_s
\hat{\bar{V_c}}+\frac{g^\prime}{3}\hat{V^\prime})}\hat{d}^{\prime
c}_{\alpha L} \right]. \label{lg3}
\end{eqnarray}
Finally, the last term can be written as
\begin{eqnarray}
{\mathcal L}_{scalar} &=& \int d^{4}\theta\;\left\{\, \hat{ \bar{
\chi}}e^{2[g\hat{V}+g^{\prime} \left( - \frac{1}{3}\right)
\hat{V}^{\prime}]} \hat{ \chi} + \hat{ \bar{
\rho}}e^{2[g\hat{V}+g^{\prime} \left( \frac{2}{3}\right)
\hat{V}^{\prime}]} \hat{ \rho} + \hat{ \bar{ \chi}}^{\prime}
e^{2[g\hat{ \bar{V}}+g^{\prime} \left( \frac{1}{3}\right)
\hat{V}^{\prime}]}
\hat{ \chi}^{\prime}\right. \nonumber \\
&&+ \left. \, \hat{ \bar{ \rho}}^{\prime} e^{2[g\hat{
\bar{V}}+g^{\prime} \left( - \frac{2}{3}\right) \hat{V}^{\prime}]}
\hat{ \rho}^{\prime} \right\} +\left( \int d^2 \theta W +
H.c\right) \! \hspace{2mm} \label{esc}
\end{eqnarray}
 with \begin{equation}
W= \frac{W_{2}}{2}+ \frac{W_{3}}{3},
\end{equation}
where
\begin{eqnarray}
W_{2}&=& \mu_{0a}\hat{L}_{aL} \hat{ \chi}^{\prime}+ \mu_{ \chi}
\hat{ \chi} \hat{ \chi}^{\prime}+
 \mu_{ \rho} \hat{ \rho} \hat{ \rho}^{\prime},
 \label{w2}
\end{eqnarray}
and
\begin{eqnarray} W_{3}&=&
\lambda_{1ab} \hat{L}_{aL} \hat{ \rho}^{\prime} \hat{l}^{c}_{bL}+
\lambda_{2a} \epsilon \hat{L}_{aL} \hat{\chi} \hat{\rho}+
\lambda_{3ab} \epsilon \hat{L}_{aL} \hat{L}_{bL}
\hat{\rho} \nonumber \\
&&+ \kappa_{1i} \hat{Q}_{1L} \hat{\chi}^{\prime} \hat{u}^{c}_{iL}+
\kappa_{1}^{\prime} \hat{Q}_{1L} \hat{\chi}^{\prime}
\hat{u}^{\prime c}_L+ \kappa_{2i}\hat{Q}_{1L} \hat{\rho}^{\prime}
\hat{d}^{c}_{iL} \nonumber \\
&&+ \kappa^\prime_{2 \alpha}\hat{Q}_{1L} \hat{\rho}^{\prime}
\hat{d}^{\prime c}_{\alpha L} + \kappa_{3 \alpha i}
\hat{Q}_{\alpha L}\hat{\rho}\hat{u}^{c}_{iL} + \kappa_{3
\alpha}^{\prime} \hat{Q}_{\alpha L}\hat{\rho}\hat{u}^{\prime
c}_{L} \nonumber \\ &&+ \kappa_{4 \alpha i} \hat{Q}_{\alpha L}
\hat{\chi} \hat{d}^{c}_{iL} + \kappa^\prime_{4 \alpha \beta}
\hat{Q}_{\alpha L} \hat{\chi} \hat{d}^{\prime c}_{\beta L}+
\epsilon f_{1\alpha\beta\gamma}\hat{Q}_{\alpha L} \hat{Q}_{\beta
L} \hat{Q}_{\gamma L} \crn &&+ \xi_{1i \beta j} \hat{d}^{c}_{iL}
\hat{d}^{\prime c}_{\beta L} \hat{u}^{c}_{j L}+ \xi_{2i \beta }
\hat{d}^{c}_{i L} \hat{d}^{\prime c}_{\beta L} \hat{u}^{\prime
c}_{L}+ \xi_{3ijk}
\hat{d}^{c}_{iL} \hat{d}^{c}_{jL} \hat{u}^{c}_{k L} \nonumber \\
&&+ \xi_{4ij} \hat{d}^{c}_{i L} \hat{d}^{c}_{jL} \hat{u}^{\prime
c}_{L}+ \xi_{5 \alpha \beta i} \hat{d}^{\prime c}_{\alpha L}
\hat{d}^{\prime c}_{\beta L} \hat{u}^{c}_{iL} + \xi_{6 \alpha
\beta} \hat{d}^{\prime c}_{\alpha L}\hat{d}^{\prime c}_{\beta L}
\hat{u}^{\prime c}_{L} \crn &&+ \xi_{7a \alpha j}\hat{L}_{aL}
\hat{Q}_{\alpha L} \hat{d}^{c}_{jL}+ \xi_{8a\alpha
\beta}\hat{L}_{aL} \hat{Q}_{\alpha L} \hat{d}^{\prime c}_{\beta
L}. \label{w3}
\end{eqnarray}
The coefficients $\mu_{0a}, \mu_{\rho}$ and $\mu_{\chi}$ have mass
dimension, while all  coefficients in $W_{3}$ are dimensionless.

To find interactions contained in the supersymmetric Lagrangian,
we first obtain all the kinetic terms \cite{susy}:
 \begin{eqnarray}
\mathcal{L}_{kinetic} &=& (D^\mu \chi)^+ D_\mu \chi +(D^\mu
\rho)^+ D_\mu
 \rho+ (\bar{D}^\mu \widetilde{Q}_{\al L})^+\bar{D}_\mu
 \widetilde{Q}_{\al L}\nonumber \\
 &&+ (\bar{D}^\mu \chi^{\prime})^+ \bar{D}_\mu \chi^{\prime} +
 (\bar{D}^\mu \rho^{\prime})^+ \bar{D}_\mu \rho^{\prime}+
 (D^\mu \widetilde{Q}_{1L})^+D_\mu
 \widetilde{Q}_{1 L}\nonumber \\
 & &+ (D^\mu \widetilde{L}_{a L})^+
 D_\mu \widetilde{L}_{a L} +(D_1^\mu
 \widetilde{d}_{iL}^c)^+D_{1\mu} \widetilde{d}_{i
 L}^c +(D_1^\mu \widetilde{u}_{i L}^c)^+
 D_{1\mu} \widetilde{u}_{i
 L}^c\crn &&+(D_1^\mu \widetilde{u}^{\prime c}_{L})^+
 D_{1\mu} \widetilde{u}^{\prime c}_{
 L}+(D_1^\mu \widetilde{d}^{\prime c}_{\alpha L})^+
 D_{1\mu} \widetilde{d}^{\prime c}_{\alpha
 L} +i\overline{L}_{aL}
 \bar{\sigma}^\mu D_\mu L_{aL} \nonumber
 \\ &&
  +i\overline{Q}_{1L}
 \bar{\sigma}^\mu D_\mu Q_{1L}
+i\overline{Q}_{\al L}
 \bar{\sigma}^\mu \bar{D}_\mu Q_{\al L} + i \overline{l}^c_{aL}
 \bar{\sigma}^\mu D_{1\mu} l^c_{aL} \nonumber
 \crn && + i \overline{u}^c_{iL}
 \bar{\sigma}^\mu D_{1\mu} u^c_{iL} + i\overline{d}^c_{iL}
 \bar{\sigma}^\mu D_{1\mu} d^c_{iL} + i \overline{u}^{\prime c}_{ L}
 \bar{\sigma}^\mu D_{1\mu} u^{\prime c}_{ L} + i \overline{d}^{\prime c}_{\al L}
\bar{\sigma}^\mu D_{1\mu} d^{\prime c}_{\al L} \crn &&-\fr 1 4
F_{ca}^{\mu \nu}F_{c a \mu \nu}-\fr 1 4 F_a^{\mu \nu}F_{a \mu \nu}
- \fr 1 4 F^{\mu \nu}F_{\mu
 \nu} + \mathcal{L}_{gaugino}+ \mathcal{L}_{Higgsinos},
 \label{Lnosusy1} \end{eqnarray}
 where
\begin{eqnarray}
\mathcal{L}_{gaugino}&=& i\bar{\lambda}^{a}_{c}
\bar{\sigma}^{\mu}D^{c}_{\mu}\lambda^{a}_{c}+i
\bar{\lambda}^{a}_{V}
\bar{\sigma}^{\mu}D^{L}_{\mu}\lambda^{a}_{V}+i
\bar{\lambda}_{B}\bar{\sigma}^{\mu}\partial_{\mu}\lambda_{B}  \,\ , \nonumber \\
\mathcal{L}_{Higgsinos}&=&i \bar{\tilde{\rho}}
\bar{\sigma}^{\mu}D_{\mu}\tilde{\rho}+ i \bar{\tilde{\chi}}
\bar{\sigma}^{\mu}D_{\mu}\tilde{\chi}+ i
\bar{\tilde{\rho}}^{\prime}
\bar{\sigma}^{\mu}\bar{D}_{\mu}\tilde{\rho}^{\prime}+ i
\bar{\tilde{\chi}}^{\prime}
\bar{\sigma}^{\mu}\bar{D}_{\mu}\tilde{\chi}^{\prime}.
\end{eqnarray}
Here, the covariant derivatives are defined by
\begin{eqnarray}
D_\mu &=&\partial_\mu + ig T^a V_{a\mu} +ig^\prime X T^9 B_\mu,
\,\ \bar{D}_\mu = \partial_\mu - ig T^{a*} V_{a\mu} +ig^\prime X
T^9 B_\mu,
\nonumber \\
D_{1\mu} &=&\partial_\mu  +ig^\prime X T^9 B_\mu, \,\
D^{c}_{\mu}\lambda^{a}_{c}=\partial_{\mu}\lambda^{a}_{c}-g_{s}
f^{abc}g^{b}_{\mu}\lambda^{c}_{c}, \nonumber \\
D^{L}_{\mu}\lambda^{a}_{V}&=&\partial_{\mu}
\lambda^{a}_{V}-gf^{abc}V^{b}_{\mu}\lambda^{c}_{V}.
\end{eqnarray}
The relevant interactions are therefore given by \cite{susy}:
\begin{eqnarray}
\mathcal{L}_{interaction}&=& \mathcal{L}_{llV}+ \mathcal{L}_{
\tilde{l} \tilde{l}V}+ \mathcal{L}_{l \tilde{l} \tilde{V}}+
\mathcal{L}_{ \tilde{l} \tilde{l}VV}+ \mathcal{L}_{qqV}+
\mathcal{L}_{ \tilde{q} \tilde{q}V}+ \mathcal{L}_{q \tilde{q}
\tilde{V}}\crn &&+ \mathcal{L}_{ \tilde{q} \tilde{q}VV}+
\mathcal{L}_{H \tilde{H} \tilde{V}} + \mathcal{L}_{llH}+
\mathcal{L}_{l \tilde{l} \tilde{H}}+ \mathcal{L}_{l\tilde{H}H}+
\mathcal{L}_{\tilde{l} \tilde{H} \tilde{H}}\crn &&+
\mathcal{L}_{qqH}+ \mathcal{L}_{q \tilde{q} \tilde{H}}+
\mathcal{L}_{lq \tilde{q}}+ \mathcal{L}_{qq \tilde{q}}+
\mathcal{L}_{qq \tilde{l}}+ V_{scalar}, \nonumber
\end{eqnarray}
where
\begin{eqnarray}
\mathcal{L}_{llV}&=&-\frac{g}{2}
\bar{L}\bar{\sigma^{\mu}}\lambda^a LV^a_{\mu}-
\frac{g^{\prime}}{\sqrt{6}} \left( -\frac{1}{3}\right)
\bar{L}\bar{\sigma^{\mu}}LB_{\mu}-
\frac{g^{\prime}}{\sqrt{6}}\bar{l}^{c}\bar{\sigma^{\mu}}l^{c}B_{\mu},
\nonumber \\
\mathcal{L}_{ \tilde{l} \tilde{l}V}&=& \frac{ig}{2}\left[
\partial^{\mu}\bar{\tilde{L}}\lambda^a
\tilde{L}- \bar{\tilde{L}}\lambda^a\partial^{\mu}
\tilde{L} \right]V^{a}_{\mu}\crn &&+
\frac{ig^{\prime}}{\sqrt{6}}\left[
-\frac{1}{3}\left(\partial^{\mu}\bar{\tilde{L}}\tilde{L}-
\bar{\tilde{L}}\partial^{\mu} \tilde{L}\right)+
\left(\partial^{\mu}\bar{\tilde{l}}^{c}\tilde{l}^{c}-
\bar{\tilde{l}}^{c}\partial^{\mu} \tilde{l}^{c}
\right)\right]B_{\mu}, \nonumber \\
\mathcal{L}_{l \tilde{l} \tilde{V}}&=&- \frac{ig}{\sqrt{2}} \left(
\bar{L}\lambda^a\tilde{L}\bar{\lambda}^a_{V}-
\bar{\tilde{L}}\lambda^aL\lambda^a_{V}\right)\crn &&-
ig^{\prime}\sqrt{\frac{1}{3}}\left[
-\frac{1}{3}\left(\bar{L}\tilde{L}\bar{\lambda}_{B}-
\bar{\tilde{L}}L\lambda_{B}\right)+
\left(\bar{l}^{c}\tilde{l}^{c}\bar{\lambda}_{B}-
\bar{\tilde{l}}^{c}l^{c}\lambda_{B}\right) \right], \nonumber \\
\mathcal{L}_{ \tilde{l} \tilde{l}VV}&=&\frac{1}{4} \left[
g^{2}V_{\mu}^aV^{b \mu}\bar{\tilde{L}} \lambda^{a}\lambda^{b}
\tilde{L}+ \frac{2}{27}g^{\prime
2}V^{\mu}V_{\mu}\bar{\tilde{L}}\tilde{L}- \fr 2 3
\sqrt{\frac{2}{3}}gg^{\prime}V^{a}_{
\mu}B^{\mu}\bar{\tilde{L}}\la^a\tilde{L}
\right]\crn &&+ \frac{g^{\prime
2}}{6}B^{\mu}B_{\mu}\bar{\tilde{l}}^{c}\tilde{l}^{c},
\nonumber \\
\mathcal{L}_{qqV}&=& -\frac{g_{s}}{2}\left(
\bar{Q}_{i}\bar\sigma^{\mu}\lambda^a Q_{i}-
\bar{u}^{c}_i\bar\sigma^{\mu}\lambda^{*a} u^{c}_i-
\bar{d}^{c}_i\bar\sigma^{\mu}\lambda^{*a} d^{c}_i- \bar{u}^{\prime
c}\bar\sigma^{\mu}\lambda^{*a} u^{\prime c}\right.\crn &&\left.-
\bar{d}^{\prime c}_{\beta}\bar\sigma^{\mu}\lambda^{*a} d^{\prime
c}_{\beta}\right)g^a_{\mu} - \frac{g}{2} \left(
\bar{Q}_{1}\bar\sigma^{\mu}\lambda^a Q_{1}- \bar{Q}_{
\alpha}\bar\sigma^{\mu}\lambda^{*a} Q_{ \alpha}\right)V^a_{\mu} -
\frac{g^{ \prime}}{\sqrt{6}}\nonumber \\
&&\times \left[ \frac{1}{3} \bar{Q}_{1}\bar\sigma^{\mu} Q_{1}-
\frac{2}{3} \bar{u}^{c}_i\bar\sigma^{\mu} u^{c}_i+ \frac{1}{3}
\bar{d}^{c}_i\bar\sigma^{\mu} d^{c}_i- \frac{2}{3} \bar{u}^{\prime
c}\bar\sigma^{\mu} u^{\prime c}+ \frac{1}{3} \bar{d}^{\prime
c}_{\beta}\bar\sigma^{\mu} d^{\prime c}_{\beta}
\right]B_{\mu}, \nonumber \\
\mathcal{L}_{ \tilde{q} \tilde{q}V}&=& \frac{ig_{s}}{2} \left(
\partial^{\mu}\bar{\tilde{Q}}_{i}\lambda^a\tilde{Q}_{i}-
\bar{\tilde{Q}}_{i}\lambda^a\partial^{\mu} \tilde{Q}_{i}-
\partial^{\mu}\bar{\tilde{u}}^{c}_i\lambda^{*a}
\tilde{u}^{c}_i\right. \crn &&+ \left.
\bar{\tilde{u}}^{c}_i\lambda^{*a}\partial^{\mu} \tilde{u}^{c}_i-
\partial^{\mu}\bar{\tilde{d}}^{c}_i\lambda^{*a}\tilde{d}^{c}_i+
\bar{\tilde{d}}^{c}_i\lambda^{*a}\partial^{\mu} \tilde{d}^{c}_i
\right.
\nonumber \\
&&- \left. \partial^{\mu}\bar{\tilde{u}}^{\prime
c}\lambda^{*a}\tilde{u}^{\prime c} + \bar{\tilde{u}}^{\prime
c}\lambda^{*a}\partial^{\mu} \tilde{u}^{\prime c}-
\partial^{\mu}
\bar{\tilde{d}}^{\prime c}_{\beta}\lambda^{*a}\tilde{d}^{\prime
c}_{\beta}+ \bar{\tilde{d}}^{\prime c}_{\beta}\lambda^{*a}
\partial^{\mu} \tilde{d}^{\prime c}_{\beta}\right)
g^a_{\mu}  \nonumber \\
&&+ \frac{ig}{2} \left(\partial^{\mu}\bar{\tilde{Q}}_{1}
\lambda^a\tilde{Q}_{1}- \bar{\tilde{Q}}_{1}\lambda^a\partial^{\mu}
\tilde{Q}_{1}-\partial^{\mu}\bar{\tilde{Q}}_{ \alpha}\lambda^{*a}
\tilde{Q}_{ \alpha}+ \bar{\tilde{Q}}_{
\alpha}\lambda^{*a}\partial^{\mu}
\tilde{Q}_{ \alpha}\right)V^a_{\mu} \nonumber \\
&&+ \frac{ig^{ \prime}}{\sqrt{6}} \left[ \frac{1}{3} \left(
\partial^{\mu}\bar{\tilde{Q}}_{1}\tilde{Q}_{1} - \bar{\tilde{Q}}_{1}
\partial^{\mu} \tilde{Q}_{1}\right)\right.
\crn &&- \left. \frac{2}{3}
\left( \partial^{\mu}\bar{\tilde{u}}^{c}_i
\tilde{u}^{c}_i - \bar{\tilde{u}}^{c}_i
\partial^{\mu} \tilde{u}^{c}_i\right)- \frac{2}{3} \left(
\partial^{\mu}\bar{\tilde{u}}^{\prime c}\tilde{u}^{\prime
c} -
\bar{\tilde{u}}^{\prime c} \partial^{\mu} \tilde{u}^{\prime c}\right)
 \right. \nonumber \\
&&+ \left. \frac{1}{3}
\left(\partial^{\mu}\bar{\tilde{d}}^{c}_i\tilde{d}^{c}_i -
\bar{\tilde{d}}^{c}_i
\partial^{\mu} \tilde{d}^{c}_i\right)+ \frac{1}{3} \left(
\partial^{\mu}\bar{\tilde{d}}^{\prime c}_{\beta}
\tilde{d}^{\prime
c}_{\beta} - \bar{\tilde{d}}^{\prime c}_{\beta} \partial^{\mu}
\tilde{d}^{\prime c}_{\beta}\right) \right] B_{\mu}, \nonumber \\
\mathcal{L}_{q \tilde{q} \tilde{V}}&=& \frac{-ig_{s}}{ \sqrt{2}}
\left[ \left( \bar{Q}_{i}\lambda^a\tilde{Q}_{i}-
\bar{u}^{c}_i\lambda^{*a}\tilde{u}^{c}_i-
\bar{d}^{c}_i\lambda^{*a}\tilde{d}^{c}_i- \bar{u}^{\prime
c}\lambda^{*a}\tilde{u}^{\prime c}- \bar{d}^{\prime
c}_{\beta}\lambda^{*a}\tilde{d}^{\prime c}_{\beta}\right)
\bar{\lambda}^a_{c}
\right. \nonumber \\
&&- \left. \left( \bar{\tilde{Q}}_{i}\lambda^aQ_{i}-
\bar{\tilde{u}}^{c}_i\lambda^{*a}u^{c}_i-
\bar{\tilde{d}}^{c}_i\lambda^{*a}d^{c}_i- \bar{\tilde{u}}^{\prime
c}\lambda^{*a}u^{\prime c}- \bar{\tilde{d}}^{\prime
c}_{\beta}\lambda^{*a}d^{\prime c}_{\beta}\right)
\lambda^a_{c} \right] \nonumber \\
&&- \frac{ig}{ \sqrt{2}} \left[ \left(
\bar{Q}_1\lambda^a\tilde{Q}_1-
\bar{Q}_{\alpha}\lambda^{*a}\tilde{Q}_{\alpha}\right)
\bar{\lambda}^a_{V}- \left( \bar{\tilde{Q}}_1\lambda^aQ_1-
\bar{\tilde{Q}}_{\alpha}\lambda^{*a}Q_{\alpha}\right)
\lambda^a_{V} \right]
\nonumber \\
&&- \frac{ig'}{\sqrt{3}} \left[ \left( \frac{1}{3}
\bar{Q}_1\tilde{Q}_1- \frac{2}{3} \bar{u}^{c}_i\tilde{u}^{c}_i+
\frac{1}{3} \bar{d}^{c}_i\tilde{d}^{c}_i- \frac{2}{3}
\bar{u}^{\prime c}\tilde{u}^{\prime c}+ \frac{1}{3}
\bar{d}^{\prime c}_{\beta} \tilde{d}^{\prime c}_{\beta} \right)
\bar{\lambda}_{B}
\right. \nonumber \\
&&- \left. \left( \frac{1}{3}\bar{\tilde{Q}}_1Q_1- \frac{2}{3}
\bar{\tilde{u}}^{c}_iu^{c}_i+ \frac{1}{3}
\bar{\tilde{d}}^{c}_id^{c}_i- \frac{2}{3} \bar{\tilde{u}}^{\prime
c}u^{\prime c}+ \frac{1}{3} \bar{\tilde{d}}^{\prime
c}_{\beta}d^{\prime c}_{\beta} \right) \lambda_B \right],
\nonumber \\
\mathcal{L}_{ \tilde{q} \tilde{q}VV}&=&\frac{1}{4} \left \{
g_{s}^{2} \left[ \bar{\tilde{Q}}_{i}
\lambda^{a}\lambda^{b}\tilde{Q}_{i}+
\bar{\tilde{u}}^{c}_i\lambda^{*a}\lambda^{*b}\tilde{u}^{c}_i+
\bar{\tilde{d}}^{c}_i\lambda^{*a}\lambda^{*b}\tilde{d}^{c}_i\right.
\right. \crn &&+\left. \left. \bar{\tilde{u}}^{\prime
c}\lambda^{*a}\lambda^{*b}\tilde{u}^{\prime c}+
\bar{\tilde{d}}^{\prime
c}_{\beta}\lambda^{*a}\lambda^{*b}\tilde{d}^{\prime c}_{\beta})
\right]g^{a}_{\mu}g^{b \mu} \right. \nonumber \\
&&+ \left. g^{2} \left[ \bar{\tilde{Q}}_{\alpha}
\lambda^{*a}\lambda^{*b}\tilde{Q}_{\alpha}+ \bar{\tilde{Q}}_{1}
\lambda^{a}\lambda^{b}\tilde{Q}_{1} \right] V^{a}_{\mu}V^{b
\mu}\right. \crn &&+ \left. \fr 2 3 g^{\prime 2} \left[ \left(
\frac{1}{3} \right)^{2}\bar{\tilde{Q}}_{1}\tilde{Q}_{1}+ \left(
\frac{-2}{3} \right)^{2}\bar{\tilde{u}^{c}}_{i}\tilde{u^{c}}_{i}+
\left( \frac{1}{3}
\right)^{2}\bar{\tilde{d}^{c}}_{i}\tilde{d^{c}}_{i}
\right. \right. \nonumber \\
&&+ \left. \left. \left( \frac{-2}{3}
\right)^{2}\bar{\tilde{u}^{\prime c}}\tilde{u^{\prime c}}+ \left(
\frac{1}{3} \right)^{2}\bar{\tilde{d}^{\prime
c}}_{\alpha}\tilde{d^{\prime c}}_{\alpha}
\right]B_{\mu}B^{\mu}\right. \crn && +  \left. 2g_{s}g
\left[\bar{\tilde{Q}}_{1} \lambda^{a}\lambda^{b}\tilde{Q}_{1}-
\bar{\tilde{Q}}_{\alpha} \lambda^{a}\lambda^{*b}\tilde{Q}_{\alpha}
\right] g^{a}_{\mu}V^{b \mu} +  \frac{2}{3}\sqrt{\frac{2}{3}}
gg^{\prime} \bar{\tilde{Q}}_{1}\lambda^{a}\tilde{Q}_{1}
V^{a}_{\mu}B^{\mu}\right. \crn &&+\left. 2\sqrt{\fr{2}{3}}g_{s}g'
\left[ \left( \frac{1}{3}\right) \bar{\tilde{Q}}_{1}
\lambda^{a}\tilde{Q}_{1}+ \left(\frac{-2}{3}
\right)\bar{\tilde{u}}^{c}_{i}\lambda^{a}\tilde{u^{c}}_{i}+ \left(
\frac{1}{3} \right)\bar{\tilde{d}}^{c}_{i}
\lambda^{a}\tilde{d^{c}}_{i} \right. \right. \nonumber \\
&&+ \left. \left. \left( \frac{-2}{3}
\right)\bar{\tilde{u}}^{\prime c} \lambda^{a}\tilde{u}^{\prime c}+
\left( \frac{1}{3} \right)\bar{\tilde{d}}^{\prime c}_{\alpha}
\lambda^{a}\tilde{d}^{\prime c}_{\alpha} \right]g^{a}_{\mu}B^{\mu}
\right \} , \nonumber \\
\mathcal{L}_{H \tilde{H} \tilde{V}}&=&- \frac{ig}{ \sqrt{2}}
\left[ \bar{\tilde{\rho}}\lambda^a\rho\bar{\lambda}^a_{V} -
\bar{\rho}\lambda^a\tilde{\rho}\lambda^a_{V}+
\bar{\tilde{\chi}}\lambda^a\chi\bar{\lambda}^a_{V} -
\bar{\chi}\lambda^a\tilde{\chi}\lambda^a_{V} -
\bar{\tilde{\rho}}^{\prime}\lambda^{*
a}\rho^{\prime}\bar{\lambda}^a_{V} \right. \crn &&+ \left.
\bar{\rho}^{\prime}\lambda^{*
a}\tilde{\rho}^{\prime}\lambda^a_{V}-
\bar{\tilde{\chi}}^{\prime}\lambda^{*
a}\chi^{\prime}\bar{\lambda}^a_{V}
 + \bar{\chi}^{\prime}\lambda^{*
a}\tilde{\chi}^{\prime}\lambda^a_{A}\right]\crn &&- \frac{ig^{
\prime}}{ \sqrt{3}} \left[ - \frac{1}{3} \left(\bar{\tilde{\chi}}
\chi \bar{\lambda}_{B}- \bar{\chi}\tilde{\chi}\lambda_{B}\right) +
\frac{1}{3} \left(\bar{\tilde{\chi}}^{\prime} \chi^{\prime}
\bar{\lambda}_{B}-
\bar{\chi}^{\prime}\tilde{\chi}^{
\prime}\lambda_{B}\right) \right. \nonumber \\
&&\left.+\frac{2}{3} \left(\bar{\tilde{\rho}} \rho
\bar{\lambda}_{B}- \bar{\rho}\tilde{\rho}\lambda_{B}\right) -
\frac{2}{3} \left(\bar{\tilde{\rho}}^{\prime} \rho^{\prime}
\bar{\lambda}_{B}-
\bar{\rho}^{\prime}\tilde{\rho}^{\prime}\lambda_{B}\right)
\right], \nonumber \\
\mathcal{L}_{l \tilde{l} \tilde{H}}&=&- \frac{ \lambda_{1}}{3}
\left( L \tilde{\rho}^{\prime} \tilde{l}^{c}+ \tilde{L}
\tilde{\rho}^{\prime}l^{c} \right)- \frac{ \lambda_{4}}{3} \left(
L \tilde{ \rho} \tilde{L}+\tilde{L} \tilde{ \rho}L \right), \crn
\mathcal{L}_{llH}&=&- \frac{ \lambda_{1}}{3} Ll^{c} \rho^{\prime}-
\frac{ \lambda_{4}}{3} LL \rho
, \nonumber \\
\mathcal{L}_{l\tilde{H}H}&=&- \frac{ \lambda_{2}}{3} \left(
L \tilde{\chi} \rho + \tilde{\rho}L \chi \right); \,\
\mathcal{L}_{\tilde{l}\tilde{H} \tilde{H}}=- \frac{ \lambda_{2}}{3}
\tilde{\chi} \tilde{\rho} \tilde{L}, \nonumber \\
\mathcal{L}_{qqH}&=&- \frac{1}{3}\left[ \kappa_{3}u^{c}Q_{\alpha}
\rho+ \kappa^{\prime}_{3}u^{\prime c}Q_{\alpha} \rho + \kappa_{4
\alpha i}d^{c}_{i}Q_{\alpha} \chi + \kappa^{\prime}_{4 \alpha
\beta}d^{\prime c}_{\beta}Q_{\alpha} \chi \right.  \crn &&\left.+
\kappa_{1i}u^{c}_{i}Q_{1} \chi^{\prime}
+\kappa^{\prime}_{1}u^{\prime c}Q_{1} \chi^{\prime}+
\kappa_{2i}d^{c}_{i}Q_{1} \rho^{\prime}+
\kappa^{\prime}_{2 \beta}d^{\prime c}_{
\beta}Q_{1} \rho^{\prime} \right], \nonumber \\
\mathcal{L}_{q \tilde{q} \tilde{H}}&=&- \frac{1}{3}\left[
\kappa_{3 \alpha i}\left(Q_{\alpha} \tilde{\rho}
\tilde{u}^{c}_{i}+ \tilde{\rho}u^{c}_{i}
\tilde{Q}_{\alpha}\right)+ \kappa^{\prime}_{3
\alpha}\left(Q_{\alpha} \tilde{\rho} \tilde{u}^{\prime c}+
\tilde{\rho}u^{\prime c} \tilde{Q}_{\alpha}\right)\right.\crn &&+
\kappa_{4 \alpha i}\left(Q_{\alpha} \tilde{\chi}
\tilde{d}^{c}_{i}+ \tilde{\chi}d^{c}_{i} \tilde{Q}_{\alpha}\right)
+\kappa^{\prime}_{4 \alpha \beta}\left(Q_{\alpha} \tilde{\chi}
\tilde{d}^{\prime c}_{\beta}+ \tilde{\chi}d^{\prime c}_{\beta}
\tilde{Q}_{\alpha}\right)\crn &&+ \kappa_{1i}\left(Q_{1}
\tilde{\chi}^{\prime} \tilde{u}^{c}_{i}+
\tilde{\chi}^{\prime}u^{c}_{i} \tilde{Q}_{1}\right)
+\kappa^{\prime}_{1}\left(Q_{1} \tilde{\chi}^{\prime}
\tilde{u}^{\prime c}+ \tilde{\chi}^{\prime}u^{\prime c}
\tilde{Q}_{1}\right)\crn &&\left.+ \kappa_{2i}\left(Q_{1}
\tilde{\rho}^{\prime} \tilde{d}^{c}_{i}+
\tilde{\rho}^{\prime}d^{c}_{i} \tilde{Q}_{1}\right)+
\kappa^{\prime}_{2 \beta}\left(Q_{1} \tilde{\rho}^{\prime}
\tilde{d}^{\prime c}_{\beta}+
\tilde{\rho}^{\prime}d^{\prime c}_{\beta}
\tilde{Q}_{1}\right)\right], \nonumber \\
\mathcal{L}_{lq \tilde{q}}&=&- \frac{\xi_{7}}{3} \left(
LQ_{\alpha} \tilde{d}^{c}+d^{c}L \tilde{Q}_{\alpha} \right)-
\frac{\xi_{8}}{3} \left( LQ_{\alpha} \tilde{d}^{\prime
c}+d^{\prime c}L \tilde{Q}_{\alpha} \right),\crn
\mathcal{L}_{ \tilde{l}qq}&=&- \frac{\xi_{7}}{3}Q_{\alpha}d^{c}
\tilde{L}-
\frac{\xi_{8}}{3}Q_{\alpha}d^{\prime c} \tilde{L}, \nonumber \\
\mathcal{L}_{qq \tilde{q}}&=&- \frac{1}{3}\left[ f_{1}QQ
\tilde{Q}+ \xi_{1}\left(d^{c}d^{\prime c} \tilde{u}^{c}+
\tilde{d}^{c}d^{\prime c}u^{c}+d^{c}\tilde{d}^{\prime
c}u^{c}\right)\right.\crn &&\left.+ \xi_{2}\left(d^{c}d^{\prime c}
\tilde{u}^{\prime c}+ \tilde{d}^{c}d^{\prime c}u^{\prime
c}+d^{c}\tilde{d}^{\prime c}u^{\prime c}\right)\right. \nonumber \\
&&\left.+ \xi_{3}\left(d^{c}d^{c} \tilde{u}^{c}+
\tilde{d}^{c}d^{c}u^{c}+d^{c}\tilde{d}^{c}u^{c}\right)\right.\crn
&&\left.+ \xi_{4}\left(d^{c}d^{c} \tilde{u}^{\prime c}+
\tilde{d}^{c}d^{c}u^{\prime c}+ d^{c}\tilde{d}^{c}u^{\prime
c}\right)\right. \nonumber \\ &&\left.+ \xi_{5}\left(d^{\prime
c}d^{\prime c} \tilde{u}^{c}+ \tilde{d}^{\prime c}d^{\prime
c}u^{c}+d^{\prime c}\tilde{d}^{\prime c}u^{c}\right)\right.\crn
&&\left.+ \xi_{6}\left(d^{\prime c}d^{\prime c} \tilde{u}^{\prime
c}+ \tilde{d}^{\prime c}d^{\prime c}u^{\prime c}+ d^{\prime
c}\tilde{d}^{\prime c}u^{\prime c}\right)\right].
 \label{lepbos}
\end{eqnarray}

The scalar potential $V_{scalar}$ has a form
 \begin{eqnarray}
 V_{scalar} &=& V_F+V_D.
 \end{eqnarray}
To find $V_F$ and $V_D$, using firstly  the Euler-Lagrangian
equations for the auxiliary fields, we obtain
\begin{eqnarray}
F_{\rho}=-  \frac{\mu_{\rho}}{2}\rho^{\prime \dagger},\hs
F_{\chi}=- \frac{\mu_{\chi}}{2}\chi^{\prime \dagger},\hs
F_{\rho^{\prime}}=-  \frac{\mu_{\rho}}{2}\rho^{\dagger},\hs
F_{\chi^{\prime}}=- \frac{\mu_{\chi}}{2}\chi^{\dagger},
\end{eqnarray}
 and
\begin{eqnarray}
D^{a}&=&- \frac{g}{2} \left[ \rho^{\dagger}\lambda^a\rho+
\chi^{\dagger}\lambda^a\chi- \rho^{\prime \dagger}\lambda^{*
a}\rho^{\prime}-
\chi^{\prime \dagger}\lambda^{* a}\chi^{\prime} \right], \nonumber \\
D&=&- \fr{g^\prime}{\sqrt{6}} \left[  - \frac{1}{3} \chi^{\dagger}
\chi + \frac{1}{3} \chi^{\prime \dagger} \chi^{\prime}+
\frac{2}{3} \rho^{\dagger}\rho - \frac{2}{3} \rho^{\prime
\dagger}\rho^{\prime}
 \right].
\end{eqnarray}
The scalar potential is therefore given by
\begin{equation}
V_{scalar}= \frac{1}{2} \left( D^{a}D^{a}+DD \right)+ \left|F_\chi
\right|^2+\left|F_\rho \right|^2+\left|F_\chi^\prime
\right|^2+\left|F_\rho^\prime \right|^2. \label{part1}
\end{equation}

\subsection{The soft term}

With the help of~\cite{gig}, the most general soft
supersymmetry-breaking terms, which do not induce quadratic
divergences can be obtained.  Such terms, in general, can be
categorized as follows: (i) A scalar field $A$ with the mass term
\be -m^{2} A^{\dagger}A;\ee (ii) A gaugino $\lambda$ with the mass
terms \be - \fr 1 2 (M_{ \lambda} \lambda^{a} \lambda^{a}+ H.c.);
\ee (iii) Finally, trilinear scalar couplings have the forms
\begin{equation}
\epsilon^{ijk}A_{i}A_{j}A_{k}+ H.c.
\end{equation}

In this model, the soft terms are given by
\begin{equation}
\mathcal{L}_{soft}= \mathcal{L}_{GMT}+\mathcal{L}_{scalar}^{soft}
+ \mathcal{L}_{SMT},
\end{equation}
where \bea \mathcal{L}^{soft}_{scalar}&=& -m^2_{ \rho}\rho^{
\dagger}\rho- m^2_{ \chi}\chi^{ \dagger}\chi -
m^2_{\rho^{\prime}}\rho^{\prime \dagger}\rho^{\prime}-
m^2_{\chi^{\prime}}\chi^{\prime \dagger}\chi^{\prime},
\label{potencial} \eea and \bea -\mathcal{L}_{SMT}&=&
m^2_{aL}\widetilde{L}_{aL}^\dagger \widetilde{L}_{aL}+
m^2_{la}\widetilde{l}_{aL}^{c \dagger} \widetilde{l}_{aL}^c
+m^2_{Q1L}\widetilde{Q}_{1L}^\dagger \widetilde{Q}_{1
L}+m^2_{Q\alpha L}\widetilde{Q}_{\alpha L}^\dagger
\widetilde{Q}_{\alpha L}\crn &&+ m^2_{u_i}\widetilde{u}_{iL}^{c
\dagger} \widetilde{u}_{iL}^c + m^2_{d_i}\widetilde{d}_{iL}^{c
\dagger} \widetilde{d}_{iL}^c + m^2_{u^\prime
}\widetilde{u^\prime}_{L}^{c \dagger} \widetilde{u^\prime}_{L}^c+
m^2_{d^\prime }\widetilde{d^\prime}_{L}^{c \dagger}
\widetilde{d^\prime}_{L}^c + M^{\prime 2}_a
\chi^\dagger\widetilde{L}_{aL}\crn
&&+\varepsilon_{1ab}\widetilde{L}_{aL}\rho ^\prime
\widetilde{l}_{Lb}^c+\varepsilon_{2a}\epsilon\widetilde{L}_{aL}\chi\varrho+
\varepsilon_{3ab}\epsilon \widetilde{L}_{aL}\widetilde{L}_{bL}\rho
+ \varrho_{1i}\widetilde{Q}_{1L}\chi^\prime\widetilde{u}^c_{iL}
\crn &&+\varrho^\prime_{1}\widetilde{Q}_{1L}\chi^\prime\widetilde{
u^\prime}^c_{iL} + \varrho_{2 \alpha i}\widetilde{Q}_{\alpha
L}\rho^\prime\widetilde{u}^c_{iL}+\varrho_{2 \alpha
}\widetilde{Q}_{\alpha L}\rho^\prime\widetilde{u^\prime}^c_{L}+
\varrho_{3i}\widetilde{Q}_{1L}\rho^\prime\widetilde{d}^c_{iL} \crn
&&+ \varrho^\prime
_{3i}\widetilde{Q}_{1L}\rho^\prime\widetilde{d^\prime}^c_{L}
+\varrho_{4\alpha i}\widetilde{Q}_{\alpha
L}\chi\widetilde{d}^c_{iL}+\varrho^\prime_{4\alpha \beta
}\widetilde{Q}_{\alpha L}\chi\widetilde{d^\prime}^c_{\beta L}\crn
&& + \varrho_{5\alpha\beta\gamma}\widetilde{Q}_{\alpha L}
\widetilde{Q}_{{\beta} L}\widetilde{Q}_{{\gamma} L}
+\kappa_{1i\beta j}\widetilde{d}_{iL}^c\widetilde{d^\prime}_{\beta
L}^c\widetilde{u}_{jL}^c+\kappa_{2i\beta
}\widetilde{d}_{iL}^c\widetilde{d^\prime}_{\beta
L}^c\widetilde{u^\prime}_{L}^c \nonumber \\ && +\kappa_{3i
jk}\widetilde{d}_{iL}^c\widetilde{d}_{j L}^c\widetilde{u}_{kL}^c
+\kappa_{4i k}\widetilde{d}_{iL}^c\widetilde{d}_{j
L}^c\widetilde{u^\prime}_{L}^c +\kappa_{5\alpha \beta
i}\widetilde{d^\prime}_{\alpha L}^c\widetilde{d^\prime}_{\beta
L}^c\widetilde{u}_{i L}^c \crn && + \kappa_{6\alpha \beta
}\widetilde{d^\prime}_{\alpha L}^c\widetilde{d^\prime}_{\beta
L}^c\widetilde{u^\prime}_{ L}^c + \kappa_{7a\alpha j
}\widetilde{L}_{a L}\widetilde{Q}_{\alpha L}\widetilde{d}_{j
L}^c+\kappa_{8a\alpha \beta }\widetilde{L}_{a
L}\widetilde{Q}_{\alpha L}\widetilde{d^\prime}_{\beta L}^c \crn
&&+ H.c.  \label{mme}\eea in order to give appropriate masses to
the sfermions.

Finally, the Lagrangian $\mathcal{L}_{GMT}$ has the same form
given in~\cite{s331r} \bea \mathcal{L}_{GMT}&=&- \frac{1}{2} [m_{
\lambda_{c}} \sum_{b=1}^{8} \left( \lambda^{b}_{c} \lambda^{b}_{c}
\right) +m_{ \lambda} \sum_{b=1}^{8} \left( \lambda^{b}_{V}
\lambda^{b}_{V} \right) +  m^{ \prime} \lambda_{B} \lambda_{B}+
H.c.].\eea This part gives masses for the superpartners of gauge
bosons.

\section{\label{gaugeboson}Gauge bosons}
In the section (\ref{sec:massspectrum}), we will prove that in
order to eliminate linear terms in the Higgs potential, one
obtains a matching condition $u/w=u'/w'$. In the following the
notation, \be t_{\theta} \equiv \fr{u}{w}=\fr{u'}{w'},
\label{ht2tan}\ee is therefore used, where $s_\theta \equiv \sin
\theta$, $t_\theta \equiv \tan \theta$, and so forth.

The mass Lagrangian for the gauge bosons can be obtained by
 \bea
2 \mathcal{L}_{mass}^{gauge}&=& \left(u,0,w \right)\left(
\frac{g}{2} \lambda^a V_{ a}^\mu- \frac{1}{3}
   \frac{g^\prime}{2} \sqrt{\frac{2}{3}}  B^\mu\right)^2
\left (u, 0, w\right)^T \crn
   & &+ \left(0,v,0\right)\left( \frac{g}{2}\la^a V_{ a}^\mu+\frac{2}{3}
\frac{g^\prime}{2} \sqrt{\frac{2}{3}} B^\mu\right)^2 \left (0, v, 0
\right)^T\crn  & &+ \left(u^\prime,  0,  w^\prime \right)
     \left( -\frac{g}{2}\lambda^{a*} V^{\mu}_a+\frac{1}{3}
 \frac{g^\prime}{2} \sqrt{\frac{2}{3}} B^\mu\right)^2
 \left ( u^\prime, 0, w^\prime\right)^T\crn & & +\left(0, v^\prime, 0
\right)\left( -\frac{g}{2}\lambda^{a*} V^{\mu }_a- \frac{2}{3}
  \frac{g^\prime}{2} \sqrt{\frac{2}{3}}
  B^\mu\right)^2
 \left ( 0,  v^\prime,  0\right)^T. \eea

Let us define the charged gauge bosons as follows \be
W'^{\pm}_\mu\equiv\fr{1}{\sqrt{2}}(V_{1\mu}\mp iV_{2\mu}),\hs
Y'^\pm_\mu \equiv \fr{1}{\sqrt{2}}(V_{6\mu}\pm V_{7\mu}).\ee The
mass matrix of the $W'_\mu$ and $Y'_\mu$ is obtained then
\begin{eqnarray}
M_{charged}^2  &=& \frac{g^2}{4}  \left(
                       \begin{array}{cc}
                         V^2+U^2& K   \\
                         K &  W^2+V^2   \\
                         \end{array}
                     \right),
\end{eqnarray}
where \bea V^2 &\equiv& v^2+v^{\prime 2},\hs W^2\equiv
w^2+w^{\prime 2},\hs U^2 \equiv u^2+u^{\prime 2}=t^2_\theta
W^2,\crn K &\equiv& uw+u^\prime w^\prime=t_\theta W^2,\hs t\equiv
g^\prime/g.\eea This matrix gives the eigenstates which are,
respectively, the SM-like $W^{\pm}$ and new gauge boson $Y^{\pm}$:
\be W_\mu = c_\theta W'_\mu-s_\theta Y'_\mu,\hs Y_\mu=s_\theta
W'_\mu +c_\theta Y'_\mu,\ee with the respective eigenvalues: \bea
m^2_{W}&=&\frac{g^2}{4}V^2, \hs
m^2_{Y}=\frac{g^2}{4}\left(V^2+U^2+W^2\right). \eea Therefore, the
$\theta$ is the mixing angle of $W'-Y'$, which is the same as in
the case of non-supersymmetric model \cite{haihiggs}. Because of
the constraint (\ref{contraint}), the mass of $W$ boson is
identified with those of the SM, that is \be
 \sqrt{v^2+v'^2}\equiv v_{\mathrm{weak}}=246\ \mathrm{GeV}.\ee

For the remaining gauge vectors $(V_3,V_8, B,V_4,V_5)$, the mass
matrix in this basis is given by \bea
 M^2_{neutral} &=& \left(
 \begin{array}{cc}
       M_{mixing}^2&0\\
  0 & M^2_{V_5}
      \end{array}
   \right),
\eea where $V_5$ is decoupled with the mass \be
 M^2_{V_5}\equiv \frac{g^2}{4}\left(W^2+U^2\right),\label{klw5}\ee
while the mixing part $M^2_{mixing}$ of $(V_3,V_8,B,V_4)$ is equal
to \bea \frac{g^2}{4}\left(
   \begin{array}{cccc}
   U^2+V^2 & \frac{1}{\sqrt{3}}\left (U^2-V^2\right)
   & -\frac{2t}{3\sqrt{6}}\left(U^2+2V^2\right)
    & K \\
    & \frac{1}{3}\left(V^2+U^2+4W^2 \right) &
\frac{\sqrt{2} t}{9}\left(2V^2+2W^2-U^2 \right)& -\frac{1}{\sqrt{3}}K \\
&  & \frac{2t^2}{27}\left( 4V^2+U^2+W^2 \right) & -\frac{4t}{3\sqrt{6}}K \\
        &  &  &
         U^2+W^2 \\
                \end{array}
  \right) \label{M4neu}
\eea
 As in the non-supersymmetric version,
it can be checked that the  matrix (\ref{M4neu}) contains two {\it
exact} eigenvalues,
  the photon $A_\mu$ and new $V'_{4\mu}\sim V_{4\mu}$,
  such as
   \bea
   M^2_\gamma &=& 0, \crn
   M^2_{V^\prime_4}&=& \frac{g^2}{4}\left(U^2+W^2 \right).
  \label{xo} \eea
Due to the fact that $V^\prime_4$ and $V_5$ gain the same mass
[cf. (\ref{xo}) and (\ref{klw5})], it is worth noting that these
boson vectors have to be combined to produce the following
physical state \cite{haihiggs}\be X^0_\mu
\equiv\fr{1}{\sqrt{2}}(V^\prime_{4\mu}-iV_{5 \mu}),\label{dnx} \ee
with the mass \be m^2_X=\fr{g^2}{4}(U^2+W^2).\ee

To look for eigenstates corresponding to eigenvalues in
(\ref{xo}), we will separate  the square mass matrix
   (\ref{M4neu}) into two parts such as
   \be
  M^2_{mixing}= M_1^2+ M_2^2,
   \ee
with \bea M^2_1=\frac{g^2}{4}\left(
     \begin{array}{cccc}
     u^2+v^2 & \frac{1}{\sqrt{3}}
     \left(u^2-v^2\right) & -\frac{2t}{3\sqrt{6}}\left(u^2+2v^2\right)
           & uw \\
          & \frac{1}{3}\left(v^2+u^2+4w^2 \right) &
 \frac{\sqrt{2} t}{9}\left(2v^2+2w^2-u^2 \right)& -\frac{1}{\sqrt{3}}uw \\
  &  & \frac{2t^2}{27}\left( 4v^2+u^2+w^2
  \right) & -\frac{4t}{3\sqrt{6}}uw \\
           &  &  &
           u^2+w^2 \\
           \end{array}
           \right) \label{M1}
           \eea
and $M^2_2$ equal to \bea \frac{g^2}{4}\left(
  \begin{array}{cccc}
   u^{\prime 2}+v^{\prime 2} & \frac{1}{\sqrt{3}}
 \left(u^{\prime 2}-v^{\prime 2}\right) &
 -\frac{2t}{3\sqrt{6}}\left(u^{\prime 2}
        +2v^{2 \prime}\right)
      & u^\prime w^\prime \\
    & \frac{1}{3}\left(v^{2 \prime}+u^{2 \prime}+4w^{2\prime }\right) &
    \frac{\sqrt{2} t}{9}\left(2v^{\prime 2}+2w^{2 \prime}-u^{2 \prime}
    \right)& -\frac{1}{\sqrt{3}}u^\prime w^\prime \\
     &  & \frac{2t^2}{27}\left( 4v^{2 \prime}+u^{2 \prime}+w^{2
     \prime }\right) & -\frac{4t}{3\sqrt{6}}u^\prime w^\prime \\
        &  &  &
             u^{2 \prime}+w^{2 \prime} \\
      \end{array}
     \right). \label{M2}
     \eea
 It is easy to realize that $M^2_1$ and $M^2_2$ have the same form
 as in  non-supersymmetric version in Ref.~\cite{haihiggs}.
 Therefore $M_1^2$ and $M_2^2$, respectively, contain the two
 eigenvalues as follows
 \bea
 \lambda_1&=& 0,\hs \la_2= \frac{g^2}{4} \left(u^2+w^2\right), \crn
\lambda'_1&=& 0,\hs \la'_2= \frac{g^2}{4} \left(u^{\prime
2}+w^{\prime 2}\right). \eea This means also that the  matrix $
M^2_{mixing}$ contains a massless eigenstate corresponding to the
photon \cite{dl} \be A_\mu^T= \frac{1}{\sqrt{18+4t^2}}\left(
        \begin{array}{cccc}
          \sqrt{3}t, & -t, & 3\sqrt{2}, & 0 \\
        \end{array}
      \right).\label{klw4}
\ee

Before seeking the eigenstate of $M^2_{V_4^\prime}$, we note that
the eigenstates according to eigenvalues $\lambda_{2}$ and
$\lambda'_{2}$ are, respectively, given by \bea V^{\prime T}_{4
\mu \lambda_{2}}&=& \frac{1}{\sqrt {1+4\La^2}}\left(
                        \begin{array}{cccc}
    \La, & \sqrt{3}\La, & 0,  & 1\\
                        \end{array}
  \right),\hs \La  \equiv \frac{2uw}{w^2-u^2}, \nonumber \\
V^{\prime T }_{4 \mu \lambda'_{2}} &=& \frac{1}{\sqrt
{1+4\La'^2}}\left(
\begin{array}{cccc}
 \La', & \sqrt{3}\La', & 0,  & 1\\
\end{array}
                      \right),\hs \La' \equiv \frac{2u^\prime
                      w^\prime}{w^{\prime2}-u^{\prime2}}.
\eea Because of the condition (\ref{ht2tan}), we get a beautiful
result
 $$V^{\prime T}_{4\mu }=
 V^{\prime T}_{4 \mu \lambda_{2}} =
 V^{\prime T}_{4 \mu \lambda'_{2}}
 =\frac{1}{\sqrt {1+4t^2_{2\theta}}}\left(
                        \begin{array}{cccc}
t_{2\theta}, & \sqrt{3}t_{2\theta}, & 0, & 1\\
                        \end{array}
                      \right).$$
These results are the same as in Ref.~\cite{haihiggs} of the
non-supersymmetric version.

The eigenvectors $A_\mu$ and $V'_{4\mu}$ can be rewritten as
follows \bea A_\mu &=& s_W V_{3 \mu} +c_W
\left(-\frac{t_W}{\sqrt{3}}V_{8 \mu}+ \sqrt{1-\frac{t^2_W}{3}}
\right)B_\mu,
\nonumber \\
V^\prime_{4\mu}&=&\frac{t_{2\theta}}{\sqrt{1+4t_{2\theta}^2}}
V_{3\mu} +\frac{\sqrt{3}t_{2\theta}}{\sqrt{1+4t_{2\theta}^2}}
V_{8\mu} +\frac{1}{\sqrt{1+4t_{2\theta}^2}} V_{4\mu}, \eea where
$s_W \equiv \sin\theta_W=\sqrt{3t^2/(18+4t^2)}$ \cite{dl}.
Further, let us define two gauge vectors
 \bea
Z_\mu&=&c_WV_{3\mu}-s_W \left(-\frac{t_W}{\sqrt{3}}V_{8 \mu}+
\sqrt{1-\frac{t^2_W}{3}} \right)B_\mu,
\nonumber \\
Z^\prime_\mu &=&\frac{t_W}{\sqrt{3}}B_{ \mu}+
\sqrt{1-\frac{t^2_W}{3}} V_{8 \mu}, \eea which are orthogonal to
$A_\mu$. To look for two last eigenstates, we will use the
argument in~\cite{haihiggs} to define \bea Z_{1\mu} &=&
c_{\theta^\prime} Z_\mu -s_{\theta^\prime}\left[t_{\theta^\prime}
\sqrt{4c_W^2-1} Z^\prime_\mu +\sqrt{1-
t^2_{\theta^\prime}\left(4c^2_W-1\right)}V_{4\mu}\right ],
\nonumber \\
Z_{1\mu}^\prime &=&\sqrt{1-
t^2_{\theta^\prime}\left(4c^2_W-1\right)}Z^\prime_{\mu}-t_{\theta^\prime}
\sqrt{4c_W^2-1} V_{4\mu}, \eea where $s_{\theta'}\equiv
t_{2\theta}/c_W/\sqrt{1+4t^2_{2\theta}}$. Then, in the base of
$\left ( A_\mu ,Z_{1\mu},Z_{1\mu}^\prime V^\prime_{4\mu}\right)$,
the squared-mass matrix $M_{neutral}^2$ becomes \be
M_{neutral}^{\prime2}=\left(\begin{array}{cccc}
                          0 & 0 & 0 & 0 \\
0 & m^2_{Z_1} & m^2_{Z_1Z_1^\prime} & 0 \\
 0 & m^2_{Z_1Z_1^\prime} & m^2_{Z_1^\prime} & 0 \\
 0 & 0 & 0& \frac{g^2}{4}\left(W^2+U^2\right)
                        \end{array}\right)
\ee with \bea m^2_{Z_1}&=&
g^2\left[\left(1+3t^2_{2\theta}\right)U^2+\left
 (1+4t^2_{2\theta}\right)V^2-t^2_{2\theta}W^2\right]
 \crn &&
 \times\left\{4\left[c_W^2+
 \left( 3-4 s^2_W\right)t^2_{2\theta}\right]
 \right\}^{-1},\nonumber \\
m^2_{Z_1Z_1^\prime}&=& g^2\left\{
\sqrt{1+4t^2_{2\theta}}\left[c_{2W} + \left(3-4 s_W^2
\right)t^2_{2\theta}\right ]U^2 - V^2 \right.\crn &&\left.
-\left(3-4 s_W^2
\right)t^2_{2\theta}W^2\right\}\left\{4\sqrt{3-4s^2_W}\left[c_W^2+\left(
3-4 s^2_W \right)t^2_{2\theta}\right]\right\}^{-1}
\label{mxing} \\
m^2_{Z_1^\prime}&=& g^2\left\{ \left[ c^2_{2W}+\left(3-4 s_{2W}^2
\right)t^2_{2\theta}\right ]U^2 +V^2\right. \crn &&
\left.+\left[4c_W^2+\left( 1+4c^2_{W}\right)\left(
3-4s^2_W\right)t^2_{2\theta}\right]W^2
 \right\}\crn &&
\times\left\{4\left(3-4s^2_W \right)\left[ c_W^2+\left(
3-4s^2_W\right)t^2_{2\theta} \right] \right\}^{-1}\nn \eea The
elements of this matrix have the same form as of
non-supersymmetric version \cite{haihiggs}, where the results are
obtained with replacement of $u, v, w$ by $U, V, W$. The last two
eigenstates and masses of the neutral gauge bosons are given as in
\cite{haihiggs}.

To finish this section, we mention again that the matrix of
neutral gauge boson mixing is separated into two terms and one of
them is the same  as in the non-supersymmetric  version. Because
of the relation among the VEVs $\om,\ \om^\prime$ and $u, \
u^\prime$, the exact diagonalization was easily performed. Here,
the gauge boson identification is the same as in
non-supersymmetric  case. This means that the imaginary part of
the non-Hermitian bilepton $X^0$ is decoupled, while its real part
has the mixing among the neutral Hermitian gauge bosons such as,
the photon, the neutral $Z$ and the extra $Z'$.

\section{\label{leptonquark}Lepton and quark sectors}
 As in Ref.\cite{marcos},  the R-charge  is chosen as follows
\begin{eqnarray}
n_{L}&=&n_{l}=n_{\rho}=n_{\rho^{\prime}}=0, \nonumber \\
n_{Q_{1}}&=&n_{u}=n_{u'}= \frac{1}{2}, \,\
n_{Q_{\alpha}}=n_{d}=n_{d'}=- \frac{1}{2}, \nonumber \\
n_{\chi}&=&1, \,\ n_{\chi^{\prime}}=-1.
\end{eqnarray}
The terms of the superpotential  with respect to this R-parity are
given by
\begin{eqnarray}
W&=&\frac{\mu_{ \chi}}{2} \hat{ \chi} \hat{ \chi}^{\prime}+
\frac{\mu_{ \rho}}{2} \hat{ \rho} \hat{ \rho}^{\prime}+
\frac{1}{3} \left[ \lambda_{1ab} \hat{L}_{aL} \hat{ \rho}^{\prime}
\hat{l}^{c}_{bL}+ \lambda_{3ab} \epsilon \hat{L}_{aL} \hat{L}_{bL}
\hat{\rho}\right. \nonumber \\ &+& \left. \kappa_{1i} \hat{Q}_{1L}
\hat{\chi}^{\prime} \hat{u}^{c}_{iL}+ \kappa_{1}^{\prime}
\hat{Q}_{1L} \hat{\chi}^{\prime} \hat{u}^{\prime c}_L +
\kappa_{2i}\hat{Q}_{1L} \hat{\rho}^{\prime} \hat{d}^{c}_{iL}+
\kappa^\prime_{2 \alpha}\hat{Q}_{1L} \hat{\rho}^{\prime}
\hat{d}^{\prime c}_{\alpha L}\right. \nonumber \\ &+& \left.
\kappa_{3 \alpha i} \hat{Q}_{\alpha L}\hat{\rho}\hat{u}^{c}_{iL} +
\kappa_{3 \alpha}^{\prime} \hat{Q}_{\alpha L}\hat{\rho}
\hat{u}^{\prime c}_L + \kappa_{4 \alpha i} \hat{Q}_{\alpha
L}\hat{\chi} \hat{d}^{c}_{iL} + \kappa^\prime_{4 \alpha \beta}
\hat{Q}_{\alpha L} \hat{\chi} \hat{d}^{\prime c}_{\beta L}
\right].
 \label{rpartsusy331rn}
\end{eqnarray}
\subsection{Charged Lepton Masses.}

From the superpotential given in Eq. (\ref{rpartsusy331rn}), it is
easy to see that the charged leptons gain mass only from the  term
\be -\frac{\lambda_{1ab}}{3}L_{aL} \rho^{\prime}l^{c}_{bL}+
H.c.\label{shklcl} \ee We therefore get mass terms \be
-\frac{\lambda_{1ab}}{3}(l_{aL}l^{c}_{bL}+
\bar{l}_{aL}\bar{l}^{c}_{bL})\frac{v^{\prime}}{\sqrt{2}}.\ee This
mass term can now be rewritten in terms of a $3 \times 3$ matrix
$X_{l}$ as follows. Defining the following two column vectors
\begin{equation}
( \psi^{+}_{l})^{T}= \left( \begin{array}{ccc} l^{c}_{1L} &
l^{c}_{2L} & l^{c}_{3L} \end{array} \right) , \,\ (
\psi^{-}_{l})^{T}= \left( \begin{array}{ccc} l_{1L} & l_{2L} &
l_{3L}
\end{array} \right),
\end{equation}
 we can rewrite our mass term as
\begin{equation}
-\mathcal{L}=(\psi^{-}_{l})^{T}X_{l}\psi^{+}_{l}+ H.c.
\end{equation}
with
\begin{equation}
X_{l}= \frac{v^{\prime}}{\sqrt{2}}\left(
\begin{array}{ccc}
  \fr{\la_{111}}{3} & \fr{\la_{112}}{3} & \fr{\la_{113}}{3} \\
  \fr{\la_{121}}{3} & \fr{\la_{122}}{3} & \fr{\la_{123}}{3} \\
  \fr{\la_{131}}{3} & \fr{\la_{132}}{3} & \fr{\la_{133}}{3}
\end{array}
\right).\nn
\end{equation}
Notice that  only VEV of $\rho^{\prime}$ is enough to give
the charged leptons  masses.

In order to get the mass eigenstates we perform the following
rotation
\begin{eqnarray}
l^{+}_{i}  &=&  V^{l}_{ij} (\psi^{+}_{l})_{j} \,\ , \nonumber \\
l^{-}_{i}  &=&  U^{l}_{ij} (\psi^{-}_{l})_{j} \,\ , \,\ i,j=1,2,3,
\label{cl1}
\end{eqnarray}
where $V^{l},U^{l}$ are unitary matrices such that
\begin{equation}
  \underbrace{{\psi^{-}_{l}}{}^T(U^{l})^T}_{{l^{-}_{i}}{}^T}
  \underbrace{(U^{l})^{*}X_{l}(V^{l})^{\dagger}}_{({\mathcal{M}}_{l})_{ij}}
  \underbrace{V^{l}\psi^{+}_{l}}_{l^{+}_{j}}.  \label{cl2}
\end{equation}
Here ${\mathcal{ M}}_{l}$ is a diagonal matrix with real
nonnegative entries
\begin{equation}
({\mathcal{M}}_{l})_{ij}=[(U^{l})^{*}X_{l}(V^{l})^{\dagger}]_{ij}
=m_{l_{i}}\delta_{ij}.
\end{equation}
The charged leptons $l_{i}$ are defined such that their absolute
masses increase with increasing $i$.

We have verified that all the charged leptons get mass which are
the same as in the usual cases \cite{ponce1}.

\subsection{Neutral Lepton Masses}
Neutrinos get masses from the term
\begin{equation}
-\fr{\lambda_{3ab}}{3}L_{aL}L_{bL}\rho+ H.c,
\end{equation}
which gives us
\begin{equation}
-\frac{\lambda_{3ab}}{3}( \nu^{c}_{aL}\nu_{bL}-
\nu_{aL}\nu^{c}_{bL}+ \overline{\nu^{c}_{aL}}\overline{\nu_{bL}}-
\overline{\nu_{aL}}\overline{\nu^{c}_{bL}}) \rho^{0}.
\end{equation}
This mass term can now be rewritten in terms of a $6 \times 6$
matrix $X_{\nu}$ by defining the following column vector
\begin{equation}
( \psi^{0}_{\nu})^{T}= \left( \begin{array}{cccccc} \nu_{1L} &
\nu_{2L} & \nu_{3L} & \nu^{c}_{1L} & \nu^{c}_{2L} & \nu^{c}_{3L}
\end{array} \right) .
\end{equation}
Now we can rewrite our mass term as
\begin{equation}
-\mathcal{\mathcal{L}}=\frac{1}{2}\left[
(\psi^{0}_{\nu})^{T}X_{\nu}\psi^{0}_{\nu}+ H.c \right],
\end{equation}
with
\begin{equation}
X_{\nu}= \frac{v}{\sqrt{2}}\left(
\begin{array}{cccccc}
  0 & 0 & 0 & 0 & G_{21} & G_{31} \\
  0 & 0 & 0 & G_{12} & 0 & G_{32} \\
  0 & 0 & 0 & G_{13} & G_{23} & 0 \\
  0 & G_{12} & G_{13} & 0 & 0 & 0 \\
  G_{21} & 0 & G_{23} & 0 & 0 & 0 \\
  G_{31} & G_{32} & 0 & 0 & 0 & 0 \\
\end{array}
\right),\nn
\end{equation}
where
\begin{equation}
G_{ab}= \frac{1}{3} \left( \lambda_{3ab}- \lambda_{3ba} \right).
\end{equation}

Due to the fact that the above matrix is symmetric, we need only
one rotation
\begin{eqnarray}
\nu_{i}  &=&  V^{\nu}_{ij} (\psi^{0}_{\nu})_{j} \,\ ,  \,\
i,j=1,2,...,6, \label{cl1}
\end{eqnarray}
where $V^{\nu}$ is an unitary matrix such that
\begin{equation}
\frac{1}{2}
  \underbrace{{\psi^{0}_{\nu}}{}^T(V^{\nu})^T}_{{\nu_{i}}{}^T}
  \underbrace{(V^{\nu})^{*}X_{\nu}(V^{\nu})^{
  \dagger}}_{({\mathcal{M}}_{\nu})_{ij}}
  \underbrace{V^{\nu}\psi^{0}_{\nu}}_{\nu_{j}}.  \label{cl2}
\end{equation}
Here ${\mathcal{M}}_{\nu}$ is a diagonal matrix with real
nonnegative entries
\begin{equation}
({ \mathcal{M}}_{\nu})_{ij}=[(V^{\nu})^{*}X_{\nu}
(V^{\nu})^{\dagger}]_{ij}=m_{\nu_{i}}\delta_{ij}.
\end{equation}
As before, the neutral leptons $\nu_{i}$ are defined such that
their absolute masses increase with increasing $i$. Due to the
fact that $G_{ab}=-G_{ba}$, the mass pattern of this sector is
$0,\ 0,$ $\ m_{\nu},\ m_{\nu},$ $\ m_{\nu},\ m_{\nu}$, where
$\sqrt{2}m_{\nu}=v\sqrt{G^{2}_{31}+G^{2}_{32}+G^{2}_{21}}$. Noting
that this mass spectrum is the same as of the non-supersymmetric
version. The quantum corrections at one loop level can be
constructed as in \cite{dls1} (see also \cite{changlong}). This
provides a realistic mass spectrum for the neutrinos.

\subsection{Masses of Up Quarks and Down Quarks}

The Yukawa couplings responsible for the masses of the up quarks
can be obtained by
\begin{eqnarray}
&-& \frac{1}{3}[ \kappa_{1i}Q_{1L} u^{c}_{iL} \chi^{\prime}+
\kappa^{\prime}_{1}Q_{1L} u^{\prime c}_L \chi^{\prime}+ \kappa_{3
\alpha i}Q_{\alpha L} u^{c}_{iL}\rho+ \kappa^{\prime}_{3
\alpha}Q_{\alpha L} u^{\prime c}_L \rho ] + H.c.
\end{eqnarray}
This mass term can now be rewritten in terms of a $4 \times 4$
matrix $X_{u}$ as follows. Defining  following two column vectors
\begin{equation}
( \psi _{u}^{+})^{T}= \left(u_{1L},\ u_{2L},\ u_{3L},\
u^{\prime}_L\right),\hs  (\psi _{u}^{-})^{T}= \left(u^{c}_{1L},\
u^{c}_{2 L},\ u^{c}_{3L},\ u^{\prime c}_L\right),
\end{equation}
 we can rewrite our mass term as
\begin{equation}
-{ \mathcal{L}}=(\psi^{-}_{u})^{T}X_{u}\psi^{+}_{u}+H.c.
\end{equation}
with
\begin{equation}
X_{u}=\frac{1}{3\sqrt{2}}\left(
\begin{array}{cccc}
 \kappa _{11}u^\prime&  \kappa _{12}u^\prime
 & \kappa _{13}u^\prime&  \kappa^{\prime}_1 u^\prime\\
  -\kappa _{321}v & -\kappa _{322}v
   & -\kappa _{323}v& -\kappa'_{32}v \\
- \kappa _{331}v & -\kappa _{332}v
& -\kappa _{333}v & -\kappa'_{33}v \\
 \kappa _{11}w^\prime&  \kappa _{12}w^\prime
 & \kappa _{13}w^\prime&  \kappa^{\prime}_1 w^\prime\\
\end{array}
\right). \label{h1}
\end{equation}
It is easily to see that the first row and the last row in the
mass matrix (\ref{h1}) are proportional. This  means that we
obtain one massless particle in the mass spectrum of the up
quarks. This problem is the same as in the non-supersymmetric
version \cite {dlhh}.

Similarly, the Yukawa couplings for the the down-quark masses are
given by
\begin{eqnarray}
- \frac{1}{3}[ \kappa_{2i}Q_{1L} d^{c}_{iL} \rho^{\prime}+
\kappa^{\prime}_{2 \beta}Q_{1L} d^{\prime c}_{\beta L}
\rho^{\prime}+ \kappa_{4 \alpha i}Q_{\alpha L} d^{c}_{iL} \chi +
\kappa^{\prime}_{4 \alpha \beta}Q_{\alpha L} d^{\prime c}_{\beta
L}\chi ] + H.c \label{h3}
\end{eqnarray}
 Defining the two column vectors
\bea (\psi _{d}^{+})^{T} &=& \left(d^{c}_{1L},\ d^{c}_{2L},\
d^{c}_{3L},\ d^{\prime c}_{2L},\ d^{\prime c}_{3L} \right),\crn
(\psi_{d}^{-})^{T} &=& \left(d_{1L},\ d_{2L},\ d_{3L},\
d^\prime_{2L},\ d^\prime _{3L}\right),\eea we can write mass term
(\ref{h3}) in the form \be -{
\mathcal{L}}=(\psi^{-}_{d})^{T}X_{d}\psi^{+}_{d}+ H.c.
\end{equation}
with
\begin{equation}
X_{d}=\frac{1}{3\sqrt{2}}\left(
\begin{array}{ccccc}
\kappa_{21}v^{\prime} & \kappa_{22}v^{\prime} &
\kappa_{23}v^{\prime} & \kappa^{\prime}_{22}v^{\prime} &
\kappa^{\prime}_{23}v^{\prime} \\
\kappa _{421}u & \kappa _{422}u & \kappa _{423}u &
\kappa^{\prime}_{422}u &
\kappa^{\prime}_{423}u \\
\kappa _{431}u & \kappa _{432}u & \kappa _{433}u &
\kappa^{\prime}_{432}u &
\kappa^{\prime}_{433}u \\
\kappa _{421}w & \kappa _{422}w & \kappa _{423}w &
\kappa^{\prime}_{422}w &
\kappa^{\prime}_{423}w \\
\kappa _{431}w & \kappa _{432}w & \kappa _{433}w &
\kappa^{\prime}_{432}w & \kappa^{\prime}_{433}w
\end{array}
\right) \label{h2}
\end{equation}
In the mass matrix (\ref{h2}), the second and the fourth rows are
proportional; the third and the last rows are proportional too.
Therefore, this matrix contains two massless particles which are
the same as in Ref. \cite{dlhh}.

The masslessness of one up-quark and two down-quarks calls for
radiative corrections. One-loop contributions can be obtained
similarly to \cite{dlhh}. We can therefore check that the quarks
get consistent masses at this level.

\section{\label{sec:massspectrum} Higgs sector}
As  mentioned above, in Eqs. (\ref{part1}) and (\ref{potencial}),
the supersymmetric Higgs potential can be written as
 \bea V_{susyeco} &\equiv & V_{scalar} + V_{soft}, \crn
 &=& \frac{\mu_{\chi}^2}{4}\left( \left|
\chi\right|^2+\left|\chi^\prime \right|^2\right)
+\frac{\mu_{\rho}^2}{4}\left( \left|
\rho\right|^2+\left|\rho^\prime \right|^2\right) \crn
&&+\frac{g^{\prime2}}{12} \left(-\frac{1}{3} \chi^{\dagger} \chi +
\frac{1}{3} \chi^{\prime \dagger} \chi^{\prime}+ \frac{2}{3}
\rho^{\dagger}\rho - \frac{2}{3} \rho^{\prime
\dagger}\rho^{\prime} \right)^2\crn & &
+\frac{g^2}{8}(\chi^\dagger_i\lambda^b_{ij}\chi_j-
\chi^{\prime\dagger}_i\lambda^{*b}_{ij}\chi^\prime_j+
\rho^\dagger_i\lambda^b_{ij}\rho_j-
\rho^{\prime\dagger}_i\lambda^{*b}_{ij}\rho^\prime_j)^2\!,\crn
&&+m^2_\rho\rho^\dagger\rho+m^2_\chi\chi^\dagger\chi
+m^2_{\rho^\prime}\rho^{\prime\dagger}\rho^\prime+
m^2_{\chi^\prime}\chi^{\prime\dagger}\chi^\prime. \label{p4} \eea

To look for mass spectrum of Higgs fields, we have to expand them
around the VEVs as \bea \chi^T&=&\left(
           \begin{array}{ccc}
            \fr{u+S_1+iA_1}{\sqrt{2}}, & \chi^{-}, & \fr{w+S_2+iA_2}{\sqrt{2}} \\
           \end{array}
         \right), \hs  \rho^T = \left(
 \begin{array}{ccc}
  \rho_1^+, & \fr{v+S_5+iA_5}{\sqrt{2}}, & \rho_2^+ \crn
          \end{array}
      \right),\label{2}\\ {\chi^\prime}^T&=&\left(
           \begin{array}{ccc}
\fr{u^\prime+S_3+iA_3}{\sqrt{2}},
& \chi^{\prime +}, & \fr{w^\prime+S_4+iA_4}{\sqrt{2}} \\
           \end{array}
         \right),\hs {\rho ^\prime}^T =\left(
                         \begin{array}{ccc}
                           \rho_1^{\prime -}, &
\fr{ v^\prime +S_6+iA_6}{\sqrt{2}}, & \rho_2^{\prime-} \\
                         \end{array}\right).\label{1}
\eea For the sake of simplicity, here we assume that the VEVs $u,\
u',\ v,\ v',\ w$ and $w'$ are real. This means that the CP
violation through the scalar exchanges is not considered in this
work.

Returning to Eq. (\ref{p4}), by  requirement of vanishing the
linear terms in fields, we get,  at the tree level approximation,
the following constraint equations \bea \mu_\chi^2+ 2 m_\chi^2&=&
-\frac{ g^{\prime 2}}{54}\left[ w^2-w^{\prime 2}+u^2-u^{\prime 2}
+ 2\left(v^{\prime 2}-v^2 \right) \right] \nonumber
\\ && -\frac{g^2}{6}\left[2\left( u^2-u^{\prime 2}+w^2-w^{\prime
2}\right) +v^{\prime 2}-v^2 \right],\\ \mu_\rho^2+2 m_\rho^2 &=&
-\frac{2 g^{2 \prime}+9g^2}{54}\left[ 2\left(v^2-v^{2 \prime}
\right) +w^{2 \prime}-w^2 +u^{ \prime 2}-u^2\right],\eea \be
m_{\chi}^2+m^2_{\chi \prime} + \mu^2_{\chi} = 0,\ee \be
m_{\rho}^2+m^2_{\rho \prime} + \mu^2_{\rho} = 0,\ee \be \left(
w^2-u^2\right)u^\prime w^\prime = \left(w^{\prime 2} -u^{\prime
2}\right)uw. \label{tuyentinh}\ee

It is noteworthy that Eq. (\ref{tuyentinh}) implies the matching
condition previously mentioned in (\ref{ht2tan}). Consequently,
the model contains a pair of Higgs triplet $\chi$ and antitriplet
$\chi^\prime$ with the VEVs in top and bottom elements governed by
the relation: $u/w =u^\prime/w^\prime$.

The squared-mass matrix derived from (\ref{p4}) can be divided
into $(6\times6)$ matrices respective to the charged, scalar and
pseudoscalar bosons. Note that there is no mixing among the scalar
and pseudoscalar bosons. We consider, first, in the case of the
pseudoscalar bosons. There are two massless particles, namely,
$A_5,A_6$.  Four others are mixing
 in the base of  ($A_1,
A_3,A_2,A_4$), their $(4\times 4)$ squared-mass matrix takes the
form: \bea M^2_{4A}=-\frac{g^2}{4}\left(
                \begin{array}{cccc}
 -w^{\prime 2} & -w^{\prime }w &
 u^{\prime }w^{\prime } & u^{\prime }w \\
  & -w^2 & uw^{\prime} & uw\\
 &  & -u^{2 \prime} & -uu^{\prime} \\
    & &  & -u^2 \\
\end{array}
\right) \label{giavohuong1} \eea

To obtain eigenvalues and eigenstates,  we change the basis to
such $(A_1^\prime, A_2^\prime, A_3^\prime, A_4^\prime)$ as \bea
A_1^\prime &=& s_\bet A_1 -c_\bet A_3 ,\hs  A_3^\prime = c_\bet
A_1 + s_\bet A_3, \crn A_2^\prime &=& s_\bet A_2 - c_\bet A_4 ,\hs
A_4^\prime = c_\bet A_2 + s_\bet  A_4, \label{candDM}\eea where
\be t_\bet \equiv \frac{w}{w^\prime}.\label{theta1}\ee Combining
this with the relation (\ref{ht2tan}), we have also \be t_\bet =
\frac{w}{w^\prime} = \frac{u}{u^\prime} \label{theta2}.\ee In the
new basis $(A^\prime_1,A^\prime_3,A^\prime_2,A^\prime_4)$, the
squared-mass matrix (\ref{giavohuong1}) can be rewritten as \bea
M^2_{4A'} &=& -\frac{g^2}{4}\left(
                     \begin{array}{cccc}
                       0 & 0 & 0 & 0 \\
    0 & -w^2-w^{\prime 2} & 0 & wu+w^{\prime }u^{\prime} \\
                       0 & 0 & 0 & 0 \\
 0 & wu+w^{\prime }u^{\prime}  & 0 &  -u^2-u^{\prime 2}\\
                     \end{array}
                   \right).
\eea We see that $A_1^\prime$ and $A_2^\prime$ are Goldstone
bosons, whereas the remaining states $A'_3$ and $A'_4$ are mixing.
Diagonalizing the later, we obtain another Goldstone boson
$\varphi_A$ and one massive state $\phi_A$ \bea \varphi_A&=&
s_\theta A_3^\prime +c_\theta A_4^\prime, \hs \phi_A= c_\theta
A_3^\prime - s_\theta A_4^\prime\label{candDM2}.\eea The mass of
$\phi_A$ is given by \bea
m^2_{\phi_A}=\frac{g^2}{4}(1+t^2_\theta)(w^2+w^{\prime 2})=
m^2_{X}. \label{hx}\eea The above equation shows that the Higgs
and gauge bosons have {\it the same } mass.

 Now we turn to the scalar sector. In this sector, six particles are mixing in
terms of an $6 \times 6 $ squared-mass matrix. In the base of
($S_1,S_2,S_3,S_4,S_5,S_6$), this matrix is given by \bea M^2_{6S}
&=& \frac{1}{2}\left(
\begin{array}{cccccc}
m_{S11}^2 & m_{S12}^2& m_{S13}^2 & m_{S14}^2 & m_{S15}^2 & m_{S16}^2\\
& m_{S22}^2 & m_{S23}^2 & m_{S24}^2 & m_{S25}^2 & m_{S26}^2 \\
 &  & m_{S33}^2 & m_{S34}^2 & m_{S35}^2 & m_{S36}^2 \\
 &  &  & m_{S44}^2 & m_{S45}^2 & m_{S46}^2 \\
 &  & &  & m_{S55}^2 & m_{S56}^2 \\
  & & &  &  & m_{S66}^2 \\
 \end{array}\right)\label{higgstrunghoa}\eea
where the matrix elements are given in the Appendix.

To study physical eigenvalues and eigenstates of
(\ref{higgstrunghoa}), we change the basis to such
$\left(S_1^\prime,S_2^\prime,S_3^\prime,S_4^\prime,S_5^\prime,S_6^\prime
\right)$ as \bea \left(
 \begin{array}{c}
 S_1 \\
 S_2 \\
 S_3 \\
 S_4\\
 S_5\\
 S_6 \\
\end{array}
\right) &=&\left(
 \begin{array}{cccccc}
s_\theta & -c_\theta
& 0 & 0 & 0 & 0 \\
 c_\theta & s_\theta & 0 & 0 & 0 & 0 \\
0 & 0 & s_\theta  & -c_\theta  & 0 & 0 \\
 0 & 0 & c_\theta & s_\theta & 0 & 0 \\
 0& 0 & 0 & 0 & \frac{v^\prime}{
 \sqrt{v^2+v'^2}} & \frac{-v}{\sqrt{v^2+v^{\prime2}}} \\
0 & 0 & 0 & 0 & \frac{v}{
\sqrt{v^2+v'^2}} & \frac{v'}{\sqrt{v^2+v^{\prime2}}}\\
\end{array}
\right)\left(
\begin{array}{c}
S_1^{\prime} \\
S_2^{\prime}\\
S_3^{\prime}\\
 S_4^{\prime} \\
 S_5^{\prime} \\
 S_6^{\prime} \\
\end{array}
 \right)
\eea

 In the new basis $(S_1^\prime, S_3^\prime,
 S_6^\prime,S_2^\prime,S_4^\prime,S_5^\prime)$, the matrix
(\ref{higgstrunghoa}) becomes
 \bea
M^2_{6S^\prime}&=& \left(
 \begin{array}{ccc}
 M^2_{3S^\prime} & 0 & 0 \\
 0 & M^2_{2S^\prime} & 0\\
 0 & 0 & 0 \\
 \end{array}
 \right).\eea
We see that the mass spectrum contains one massless particle
$S_5^\prime$. The submatrices of $(S_2^\prime,S_4^\prime)$ and
$(S_1^\prime, S_3^\prime,
 S_6^\prime)$ are decoupled and, respectively, given by \be
M^2_{2S^\prime} =\frac{g^2}{4}(1+t^2_\theta)\left(
 \begin{array}{cc}
 w'^2 &  -w w^{\prime} \\
 -w w^{\prime} & w^2 \\
 \end{array}
 \right),
 \label{22mass}\ee
\begin{flushleft}
 \be M^2_{3S^\prime}=\left(
 \begin{array}{ccc}
\fr{18 g^2 + g'^2}{54 c^2_\theta} w^2 & -\fr{18g^2 +
              g'^2}{54 c^2_\theta} w w' & \fr{g^2 (9 g^2 +
              2 g'^2)}{54 c_\theta}\sqrt{v^2 + v'^2}w\\
       & \fr{18 g^2 + g'^2}{54c^2_\theta}w'^2
      & -\fr {g^2 (9g^2 +
              2g'^2)}{54c_\theta}\sqrt{v^2 + v'^2}w' \\
        &
 & \fr{9 g^2 + 2 g'^2}{27}(v^2 + v'^2)\\
    \end{array}
  \right). \label{dfd}
 \ee
 \end{flushleft}  The matrix (\ref{22mass}) gives us one massless field
 \be \varphi_{S_{24}}=s_\beta S'_2 + c_\beta S'_4, \ee and another
 massive
 \be \phi_{S_{24}}=c_\beta S'_2 - s_\beta S'_4\ee with the mass:
 \bea
 m^2_{\phi_{S_{24}}}&=&\frac{g^2}{4}(1+t^2_\theta)(w^2+w'^2) = m_{X}^2. \eea
Let us note that $\phi_A$ and $\phi_{S_{24}}$ have the same mass,
which can be combined to become a physical neutral complex field
$H^0_X=(\phi_{S_{24}}+i\phi_A)/\sqrt{2}$ with mass equal to $m_X$
of the neutral non-Hermitian gauge boson $X^0$.

To obtain the physical fields in $M^2_{3S^\prime}$, we use the
following transformation: \bea \left(
  \begin{array}{c}
    S_{1a}^\prime \\
    S_{3a}^\prime \\
    S_{6a}^\prime \\
  \end{array}
\right)= \left(
  \begin{array}{ccc}
    c_\beta &
s_\beta & 0 \\
    -s_\beta &
c_\beta & 0 \\
    0 & 0 & 1 \\
  \end{array}
\right)\left(
         \begin{array}{c}
           S_1^\prime\\
           S_3^\prime\\
           S_6^\prime \\
         \end{array}
       \right)
 \eea
In the new basis $(S_{1a}^\prime,S_{3a}^\prime, S_{6a}^\prime)$,
the matrix (\ref{dfd}) becomes
\be M^2_{a3 S'}=\left(%
\begin{array}{ccc}
  0 & 0 & 0 \\
  0 & m^2_{33a} & -m^2_{36a} \\
  0 & -m^2_{36a}  & m^2_{66a} \\
\end{array}%
\right),\ee where \bea m^2_{33a}&=&\fr{18 g^2 +
g'^2}{54c^2_\theta}(w^2 + w'^2),\hs m^2_{66a}=\fr{9g^2 +
2g'^2}{27}(v^2 + v'^2),\crn
m^2_{36a}&=&\fr{g^2(9g^2+2g'^2)\sqrt{(v^2 + v'^2)(w^2 + w'^2)}}{54
c_\theta}.\nn \eea

The field $S_{1a}^\prime$ is physical and massless. The submatrix
of $(S_{3a}^\prime,S_{6a}^\prime)$ is decoupled, and therefore the
diagonalization   yields the eigenvalues
 \bea m^2_{\varphi_{S_{a36}}} &=& \frac{1}{2}
\left[m^2_{33a}+m^2_{66a}-
\sqrt{\left(m^2_{33a}-m^2_{66a}\right)^2 +4m^4_{36a}} \right],
\nonumber
 \\ m^2_{\phi_{S_{a36}}} &=& \frac{1}{2}
\left[m^2_{33a}+m^2_{66a}+
\sqrt{\left(m^2_{33a}-m^2_{66a}\right)^2 +4m^4_{36a}} \right],\eea
with the respective eigenstates \be \varphi_{S_{a36}}=s_\al S'_3
+c_\al S'_6,\hs \phi_{S_{a36}}=c_\al S'_3 -s_\al S'_6,\ee where
\be t_{2\al}\equiv \fr{-2m^2_{36a}}{m^2_{66a}-m^2_{33a}}.\ee

Finally, we consider the mass spectrum of charged  Higgs bosons.
In the base of $(\chi^{+},\ \chi^{+\prime},\ \rho^{+}_1,\
\rho^{+}_2,\ \rho^{+\prime}_1,\ \rho^{+\prime}_2)$, the
squared-mass matrix can be written as
 \bea
 M^2_{6charged}=\frac{g^2}{4}
\left(
 \begin{array}{cccccc}
 m^2_{\chi^-\chi^{+}} & m^2_{\chi^-\chi^{+\prime}}& uv &
vw & -uv^\prime & -v^\prime w \\
&  m^2_{\chi^{-\prime}\chi^{+\prime}} & -vu^\prime
 & -w^\prime v & v^\prime u^\prime & v^\prime w^\prime \\
 &  & m^2_{\rho^-_1\rho^+_1} &m^2_{\rho^-_1\rho_2^+} & -vv^\prime &  0 \\
  &  &
 & -m^2_{\rho^-_2\rho_2^+}&0 &-vv^\prime  \\
  &  &
 &  & m^2_{\rho_1^{-\prime}\rho_1^{+\prime}} &
 m^2_{\rho^{-\prime}_1\rho^{+\prime}_2}\\
  &  &  &  &  &
 m^2_{\rho_2^{-\prime}\rho_2^{+\prime}} \\
 \end{array}
 \right), \label{chargeh}
 \eea
where  \bea m^2_{\chi^-\chi^{+}}&=& w^{2\prime}+u^{2\prime}
+(v^2-v^{2\prime}),\hs
m^2_{\chi^-\chi^{+\prime}}=-ww^\prime-uu^\prime,\crn
m^2_{\rho^-_1\rho_2^+}&=&u^2-u^{2\prime}+v^{2\prime},\hs
m^2_{\chi^{-\prime}\chi^{+\prime}}=u^2+w^2-(v^2-v^{2\prime}),\crn
m^2_{\rho^-_1\rho_2^+}&=& uw-u^\prime w^\prime,\hs
m^2_{\rho_1^{\prime -}\rho_1^{\prime+}}=u^{\prime 2}-u^2+v^2,\hs
m^2_{\rho^{-\prime}_1
\rho^{+\prime}_2}=-m^2_{\rho^{-}_1\rho^{+}_2}, \nonumber \\
m^2_{\rho_2^-\rho_2^+}&=& w^2-w^{2\prime}+v^{2\prime},\hs
m^2_{\rho_2^{-\prime}\rho_2^{+\prime}}=w^{2\prime}-w^2 +v^2.\eea

To diagonalize the matrix (\ref{chargeh}), we choose a new basis
as follows
 \bea \left(
  \begin{array}{c}
    \chi^{+}_a \\
    \chi^{\prime +}_a \\
  \end{array}
\right) &=&  O\left(
                \begin{array}{c}
                  \chi^+ \\
                  \chi^{\prime +} \\
                \end{array}
              \right),\hs \left(
                         \begin{array}{c}
                           \rho^+_{1a} \\
                           \rho^+_{2a} \\
                         \end{array}
                       \right)=O_1\left(
                                 \begin{array}{c}
                                   \rho^+_{1} \\
                                   \rho^+_{2} \\
                                 \end{array}
                               \right),\hs \left(
                         \begin{array}{c}
                           \rho^{\prime +}_{1a} \\
                           \rho^{\prime +}_{2a} \\
                         \end{array}
                       \right)=O_1\left(
                                 \begin{array}{c}
                                   \rho^{\prime +}_{1} \\
                                   \rho^{\prime +}_{2} \\
                                 \end{array}
                               \right),\nn
\eea with \be O \equiv \left(
          \begin{array}{cc}
            s_\beta &  -c_\beta  \\
     c_\beta &  s_\beta \\
          \end{array}
        \right),\hs  O_1\equiv \left(
                          \begin{array}{cc}
                           -c_\theta & s_\theta  \\
     s_\theta & c_\theta  \\
                          \end{array}
                        \right).
\ee
 In the base of ($\chi^{+}_a , \chi^{\prime
+}_a,\rho^+_{1a},\rho^+_{2a},\rho^{\prime +}_{1a}, \rho^{\prime
+}_{2a}$), the matrix (\ref{chargeh}) becomes  \bea
M^2_{a6charged} &=& \frac{g^2}{4}\left(
  \begin{array}{cccccc}
    m^2_{a11} & m^2_{a12} & 0 & m^2_{a14} & 0
    &m^2_{a16} \\
    & m^2_{a22} & 0 & 0 & 0 & 0 \\
     & & v^{\prime 2} & 0 & -vv^\prime & 0 \\
     &  &  &
    m^2_{a44} & 0 & -vv^\prime \\
     &  &  &  & v^2 & 0 \\
     &  &  &  & & m^2_{a66} \\
  \end{array}
\right), \label{LK}\eea where \bea m^2_{a11}&=& -c_{2\beta}
\left(t^2_\gamma-1 \right)v^{\prime 2},\hs m^2_{a12}=-s_{2\beta}
\left(t^2_\gamma-1 \right)v^{\prime 2},\nonumber \\  m^2_{a22}&=&
c_{2\beta} \left(t^2_\gamma-1 \right)v^{\prime 2}+\left(
1+t^2_\beta
\right)\left( 1+t^2_\theta \right)w^{\prime 2}, \nonumber \\
m^2_{a14}&=& \sqrt{\left( 1+t^2_\beta \right)\left( 1+t^2_\theta
\right)}w^\prime v,\hs
 m^2_{a16}  = \sqrt{\left(
1+t^2_\beta \right)\left( 1+t^2_\theta \right)}w^\prime v^\prime,
\nonumber
\\ m^2_{a44}&=& \left(t^2_\beta -1 \right)\left(t^2_\theta+1
\right)w^{\prime 2}+v^{\prime 2}, \crn
m^2_{a66}&=&-\left(t^2_\beta -1 \right)\left(t^2_\theta+1
\right)w^{\prime 2}+v^{2},\hs  t_\gamma
\equiv\frac{v}{v^\prime}.\eea Since the block intersected by the
third, fifth rows and columns is decoupled, it can be diagonalized
and this yields two eigenvalues as follows \bea
m^2_{\varrho_{1}^+}&=&\frac{g^2}{4}\left(
v^2+v^{\prime 2}\right) = m_W^2,\label{qhrow}\\
m^{2}_{\varrho^{+}_2}&=&0.\eea Here the Goldstone boson
$\varrho^+_2$ and Higgs boson $\varrho^+_1$ are, respectively,
defined by \bea \varrho_{1}^+&=& c_\ga \rho_{1a}^{+} -
s_\ga\rho_{1a}^{\prime +},\hs \varrho_2^+= s_\ga\rho_{1a}^+ +
c_\ga \rho_{1a}^{\prime +}.\eea Equation (\ref{qhrow}) shows that
{\it one charged Higgs boson has the mass equal to those of $W$
boson, i. e. $m_{\varrho^{\pm}_1} = m_{W^\pm}$}, this result is in
agreement with the experimental current limit $m_{H^{\pm}} >79.3\
\mathrm{GeV},\ \mathrm{CL}=95\% $ \cite{pdg}.

The remaining part of ($\chi^{+}_a , \chi^{\prime
+}_a,\rho^+_{2a}, \rho^{\prime +}_{2a}$) is still mixing in terms
of an $4\times 4$ submatrix of (\ref{LK}). Under the constraints
(\ref{contraint}) we will split this matrix into two terms,\bea
M^2_{a4charged}= M^2_{b4charged}+M^2_{c4charged},\eea satisfying
the condition: $|M^2_{c4charged}| \ll |M^2_{b4charged}|$. In this
case, the first matrix chosen consists of elements of the
leading-order terms in $w^2$ or $w'^2$ of $M^2_{a4charged}$, while
$M^2_{c4charged}$ contains the remaining terms. The matrix
$M^2_{a4charged}$ can therefore be diagonalized  with the
contribution of $M^2_{c4charged}$ considered as a perturbation.
The eigenvalues and eigenstates are, respectively, obtained up to
the first order contributions as follows \bea m^2_{\zeta_1^+}&=&
\frac{g^2}{4}m^2_{a11},\hs \zeta_{1}^+= \chi_a^+ -
\frac{m_{a12}^2}{\widetilde{m}^2_{\zeta_2^+}}\chi_a^{\prime+}-
\frac{m_{a14}^2}{\widetilde{m}^2_{{\zeta_3^+}}}\rho_{a2}^+ -
\frac{m_{a16}^2}{\widetilde{m}^2_{\zeta_4^+}}\rho_{a2}^{\prime
+},\crn m^2_{\zeta_2^+}&=&\frac{g^2}{4}m^2_{a22},\hs \zeta_{2}^+=
\chi_a^{\prime+} +
\frac{m_{a12}^2}{\widetilde{m}^2_{\zeta_2^+}}\chi_a^{+},\crn
m^2_{\zeta_3^+}&=&\frac{g^2}{4} m^2_{a44},\hs\zeta_3^+
=\rho_{2a}^+
+\frac{m_{a14}^2}{\widetilde{m}^2_{{\zeta_3^+}}}\chi_{a}^+
-\frac{vv^\prime}{\widetilde{m}^2_{\zeta_3^+}-
\widetilde{m}^2_{\zeta_4^{+}}}\rho_{a2}^{\prime +},\crn
m^2_{\zeta_4^+}&=&\frac{g^2}{4} m^2_{a66},\hs \zeta_4^+
=\rho_{2a}^{\prime+}
+\frac{m_{a16}^2}{\widetilde{m}^2_{{\zeta_3^+}}}\chi_{a}^+
+\frac{vv^\prime}{\widetilde{m}^2_{\zeta_3^+}
-\widetilde{m}^2_{\zeta_4^+}}\rho_{a2}^{ +}.\eea  Here we have
denoted  \bea \widetilde{m}^2_{\zeta_2^+}&=&\left( 1+t^2_\beta
\right)\left( 1+t^2_\theta \right)w^{\prime 2},\nonumber \\
\widetilde{m}^2_{\zeta_3^+}&=&-\widetilde{m}^2_{\zeta_4^+}=\left(t^2_\beta
-1 \right)\left(t^2_\theta+1 \right)w^{\prime2}.\eea If we control
parameters by the condition below
 \bea
\frac{v^2}{w^{\prime2}}&=&\left(t^2_\beta -1
\right)\left(t^2_\theta+1 \right).\eea then $m^2_{\zeta_4^+}=0$.
Therefore, in the mass spectrum of charged Higgs, we have two zero
eigenvalues which correspond to four Goldstone bosons, namely
$\zeta^{\pm}_4,\ \varrho_2^{\pm}$.

Finally, let us summarize the physical fields of the scalar sector
in the model. There are eight neutral massless particles: five
pseudoscalars $A_5$, $A_6$, $A'_1$, $A'_2$, $\varphi_A$, and three
scalars $S'_5$, $\varphi_{S_{24}}$, $S'_{1a}$. There are one
complex neutral Higgs $H^0_X$ with mass equal to the mass $m_X$ of
the neutral non-Hermitian gauge boson $X$, and two massive scalars
$\varphi_{S_{a36}}$, $\phi_{S_{a36}}$. There are two charged
massless scalar fields $\varrho^{\pm}_2$ and $\varrho^{\pm}_4$,
and four massive charged bosons $\varrho^{\pm}_1$,
$\zeta^{\pm}_1$, $\zeta^{\pm}_2$, $\zeta^{\pm}_3$. The first
charged Higgs boson has a mass equal to those of $W$ boson:
$m_{\varrho^{\pm}_1} = m_W$.

\section{\label{concl} Conclusions}

In this paper, we have constructed a supersymmetric version of the
economical 3-3-1 model of Refs.~\cite{ponce,haihiggs,higgseconom}
which includes right-handed neutrinos with the minimal scalar
sector. The Higgs sector was in detail  studied: the eigenvalues
and physical states were derived. The constraint equations and the
gauge boson identification establish a relation  between the
vacuum expectation values at the top and bottom elements of the
Higgs triplet $\chi$ and its supersymmetric counterpart
$\chi^\prime$. Because of this relation, the exact diagonalization
of neutral gauge boson sector has been performed. The gauge bosons
and their associated Goldstone ones are mixing in the same way as
in non-supersymmetric version.  The matrix of neutral gauge boson
mixing is separated into two parts and one of them is those in the
non-supersymmetric version. There is similarity in the gauge boson
mixing in both supersymmetric and non-supersymmetric versions.
This is also correct in the case of gauginos.

The model contains a heavy neutral complex scalar with mass equal
to those of the neutral non-Hermitian gauge boson $X^0$ and {\it a
charged scalar with mass equal to those of the $W$ boson in the
standard model}, i.e. $m_{H_X} = m_{X},\ m_{\varrho_1^+} = m_W$.
This value is in good agreement with the present bound \cite{pdg}
 $m_{H^\pm} > 79.3$ GeV, CL=
95 \%.  We have also shown that the boson sector and the fermion
sector get masses in the same way as in the non-supersymmetric
case. We have shown that the usual quarks have masses proportional
to VEVs of the neutral scalars which belong to doublets while the
exotic quarks and new gauge bosons gain masses of order of VEVs of
the scalar of the singlets of the standard model group. The usual
quarks have masses proportional to VEVs of the neutral scalars
which belong to doublets of the standard model -
$v_{\rho},v_{\chi_{1}},v_{\rho^{\prime}}$ and
$v_{\chi^{\prime}_{1}}$, while the exotic quarks and new gauge
bosons gain masses of order of VEVs   $v_{\chi_{2}}$ and
$v_{\chi^{\prime}_{2}}$ of the scalars which break the 3-3-1
symmetry down to the standard model, in the other words, the
scalar of the singlets of the above group.  As in
non-supersymmetric version, at the tree level, one up-quark and
two down-quarks are massless. However, the one-loop correction
will give all of them consistent masses.

It is known that the economical (non-supersymmetric) 3-3-1 model
does not furnish the candidate for self-interacting dark matter.
With a larger content of the scalar sector, the supersymmetric
version is expected to have the candidate for SIDM. The
preliminary analysis leads us to conclusion that one neutral
scalar contains the properties of SIDM like stability, neutrality
and the Universe's overpopulationless. Hence, in contrast with the
non-supersymmetric version, the considered model contains the
scalar satisfying the properties of SIDM.  We will return to this
topic in the future study.

Due to the minimal content of the scalar sector, the significant
number of free parameters is reduced. In this model, the lepton
number violation exists in the neutrino and exotic quark sectors.
Certainly, the model contains very rich phenomenology and it
deserves further studies.

\section*{Acknowledgments}

D. T. H. is grateful to Aichi University of Education, Kariya,
Japan for supporting this work. H. N. L. would like to thank
Professor R. Ruffini and Members of ICRANet in Pescara for kind
help and hospitality during his visit where this work was
initialed. The work was also supported in part by National Council
for Natural Sciences of Vietnam under grant  No: 410604. M. C. R.
is grateful to Conselho Nacional de Desenvolvimento Cient\'\i fico
e Tecnol\'ogico (CNPq) under the processes 309564/2006-9 for
supporting his work.
\\[0.3cm]

\appendix
\section{The mass matrix elements for scalar  neutral Higgs bosons }
\bea
m_{S11}^2&=&\frac{g^2}{2}w^{\prime2}+\frac{1}{27}\left(18g^2+g^{\prime
2}\right)u^2,\hs
m_{S12}^2=-\frac{g^2}{2}u^{\prime}w^\prime+\frac{1}{27}\left(
18g^2+g^{2\prime}\right)uw,\nonumber \\ m_{S13}^2&=&
-\frac{g^2}{2} ww^{\prime} -
\frac{1}{27}\left(18g^2+g^{2\prime}\right)uu^\prime, \hs
m_{S14}^2=\frac{g^2}{2}u^{\prime}w-\frac{1}{27}\left(
18g^2+g^{2\prime}\right)uw^\prime \nonumber \\
m^2_{S15}&=&-\frac{g^2}{27}\left(2g^{2\prime}+9g^2 \right)uv,\hs
m^2_{S16}=\frac{g^2}{27}\left(2g^{2\prime}+9g^2
\right)uv^\prime \nonumber \\
m^2_{S22}&=&\frac{g^2}{2}u^{\prime2}+\frac{1}{27}\left(
18g^2+g^{2\prime}\right)w^2,\hs
m^2_{S23}=\frac{g^2}{2}u^{\prime}w-\frac{1}{27}\left(
18g^2+g^{2\prime}\right)u^\prime w \nonumber \\
m^2_{S24}&=&-\frac{g^2}{2}uu^\prime-\frac{1}{27}\left(
18g^2+g^{2\prime}\right)w w^\prime,\hs
m^2_{S25}=-\frac{g^2}{27}\left(
9g^2+2g^{2 \prime}\right)w v \nonumber \\
m^2_{S26}&=&\frac{g^2}{27}\left(2g^{2\prime}+9g^2 \right)w
v^\prime,\hs m^2_{S33}=\frac{1}{27}\left(g^{2\prime}+18g^2
\right)u^{\prime2}+\frac{g^2}{2}w^{ 2}\nonumber \\
m^2_{S34}&=&\frac{1}{27}\left(g^{2\prime}+18g^2 \right)u^\prime
w^\prime-\frac{g^2}{2}uw,\hs
m^2_{S35}=\frac{g^2}{27}\left(2g^{2\prime}+9g^2 \right)u^\prime v
\nonumber
\\ m^2_{36}&=&-\frac{g^2}{27}\left(2g^{2\prime}+9g^2
\right)u^\prime v^\prime,\hs
m^2_{S44}=\frac{1}{27}\left(g^{2\prime}+18g^2
\right)w^{\prime2}+\frac{g^2}{2}u^{2}\nonumber \\
m^2_{S45}&=&\frac{g^2}{27}\left(2g^{2\prime}+9g^2
\right)vw^\prime,\hs
m^2_{46}=-\frac{g^2}{27}\left(2g^{2\prime}+9g^2 \right)w^\prime
v^\prime \nonumber \\
m^2_{S55}&=&\frac{2}{27}\left(2g^{2\prime}+9g^2 \right)v^2,\hs
m^2_{S56}=- \frac{2}{27}\left(2g^{2\prime}+9g^2 \right)v v^\prime
\nonumber \\ m^2_{S66}&=&\frac{2}{27}\left(2g^{2\prime}+9g^2
\right) v^{\prime 2}.\nn\eea


\begin{thebibliography}{999}
\bibitem{superk} SuperKamiokande Collaboration, Y. Fukuda {\it et
al.}, Phys. Rev. Lett. {\bf 81}, 1158 (1998); {\bf 81}, 1562
(1998); {\bf 82}, 2644 (1999); {\bf 85}, 3999 (2000); Y. Suzuki,
Nucl. Phys. B, Proc. Suppl. {\bf 77}, 35 (1999); S. Fukuda {\it et
al.}, Phys. Rev. Lett. {\bf 86}, 5651 (2001); Y. Ashie {\it et
al.}, Phys. Rev. Lett. {\bf 93}, 101801 (2004).

\bibitem{kam} KamLAND Collaboration, K. Eguchi {\it et al.},  Phys.
Rev. Lett. {\bf 90}, 021802 (2003); T. Araki {\it et al.}, Phys.
Rev. Lett. {\bf 94}, 081801(2005).

\bibitem{sno} SNO Collaboration, Q. R. Ahmad {\it et al.}, Phys.
Rev. Lett. {\bf 89}, 011301 (2002); {\bf 89}, 011302 (2002); {\bf
92}, 181301 (2004); B. Aharmim {\it et al.}, Phys. Rev. C {\bf
72}, 055502 (2005).

\bibitem{ppf} F. Pisano, V. Pleitez, Phys. Rev.  D {\bf 46} (1992) 410;
P.H. Frampton, Phys. Rev. Lett. {\bf 69} (1992) 2889; R. Foot {\it
et al.,} Phys. Rev. D {\bf 47} (1993) 4158.

\bibitem{flt} M. Singer {\it et al.,} Phys.
Rev. D {\bf 22} (1980) 738.

 \bibitem{331rh}R. Foot, H. N. Long and Tuan A. Tran,
 Phys. Rev. D {\bf
50} (1994) 34R; J. C. Montero  {\it et al.,} Phys. Rev. D {\bf 47}
(1993) 2918; H. N. Long, Phys. Rev. D {\bf 54} (1996) 4691; H. N.
Long, Phys. Rev. D {\bf 53} (1996) 437.

\bibitem{s331r} J. C. Montero, V. Pleitez, M. C. Rodriguez,
Phys. Rev. D {\bf 70} (2004) 075004.


\bibitem{ponce}  W. A. Ponce, Y. Giraldo, L.A. Sanchez, Phys. Rev. D {\bf
67} (2003) 075001.

\bibitem{haihiggs} P. V. Dong, H. N. Long, D.T. Nhung, D. V.
Soa, Phys. Rev. D {\bf 73} (2006) 035004.

\bibitem{higgseconom} P. V. Dong, H. N. Long, D. V.
Soa, Phys. Rev. D {\bf 73} (2006) 075005.

\bibitem{dlhh} P. V. Dong, Tr. T. Huong, D. T. Huong and H. N. Long,
 Phys. Rev. D {\bf 74} (2006) 053003.

 \bibitem{dls1} P. V. Dong, H. N. Long, D. V. Soa, {\it Neutrino
masses in the economical 3-3-1 model}, hep-ph/0610381.

\bibitem{susy}
J. Wess,  J. Bagger, {\it Supersymmetry and Supergravity}, 2nd
edition, Princeton University Press, Princeton NJ, (1992);
 H. E. Haber, G. L. Kane, Phys. Rep. {\bf 117}, 75 (1985).

\bibitem{hrl}  D. T. Huong, M. C. Rodriguez,  H. N. Long, {\it
 Scalar sector of   Supersymmetric
 $\mbox{SU}(3)_C\otimes \mbox{SU}(3)_L \otimes \mbox{U}(1)_N$
 Model  with right-handed neutrinos}, [arXiv:
hep-ph/0508045].

\bibitem{marcos}  P. V. Dong, D. T. Huong, M. C. Rodriguez, H. N. Long,
Eur. Phys. J. C  {\bf 48}, (2006) 229, [arXiv: hep-ph/0604028 ].

\bibitem{pallong} H. N. Long, P. B. Pal, Mod. Phys. Lett.
{\bf A 13}, (1998) 2355.


\bibitem{sidm331} D. Fregolente, M. D. Tonasse, Phys. Lett. B {\bf 555}
(2003) 7; H. N. Long, N. Q. Lan, Europhys. Lett. {\bf 64} (2003)
571.

\bibitem{SIDM} V. Silveira, A. Zee, Phys. Lett. B {\bf 161} (1985) 136; D. E.
Holz, A. Zee, Phys. Lett. B {\bf 517} (2000) 239;  C.P. Burgess,
M. Pospelov, T. ter Veldhuis, Nucl. Phys. B {\bf 619} (2002) 709;
B. C. Bento, O. Bertolami, R. Rosenfeld, L. Teodoro, Phys. Rev. D
{\bf 62} (2000) 041302;  J. McDonald, Phys. Rev. Lett. {\bf 88}
(2002) 091304.


\bibitem{ss} D. N. Spergel, P. J. Steinhardt,
Phys. Rev. Lett. {\bf 84} (2000) 3760.



\bibitem{gig} L. Girardello, M. T. Grisaru, Nucl. Phys. B {\bf
194} (1982) 65.

\bibitem{dl} P. V. Dong and H. N. Long, Eur. Phys. J. C {\bf 42} (2005)
325.

\bibitem{ponce1} D. A. Gutierrez, W. A. Ponce, L. A. Sanchez,
Int. J. Mod. Phys. A {\bf 21} (2006) 2217.

\bibitem{changlong} D. Chang, H. N. Long, Phys. Rev. D {\bf 73}
(2006) 053006.

\bibitem{pdg} W. -M. Yao {\it et. al.} (Particle Data Group),
J. Phys. G: Nucl. Part. Phys. {\bf 33} (2006) 1, p. 32.

\end{thebibliography}
\end{document}